\documentclass[10pt]{article}
\usepackage[nofonts]{dgleich-common}

\usepackage[T1]{fontenc}
\RequirePackage[scaled=0.8]{helvet}%
\RequirePackage[scaled=0.75]{beramono}%
\def\lstbasicfont{\fontfamily{fvm}\selectfont}
\makeatletter  
\setboolean{@dgleich@sanssmallcaps}{true}
\makeatother

\usepackage{dgleich-math}
\usepackage{dgleich-tensor}
\newcommand{\erdosrenyi}{Erd\H{o}s-R\'enyi\xspace}

\graphicspath{{./figures/}}
\usenatbib
\usehyperref

\usepackage[most]{tcolorbox}
\definecolor{plots1}{rgb}{0.0,0.605603,0.97868}
\definecolor{plots2}{rgb}{0.888874,0.435649,0.278123}
\definecolor{plots3}{rgb}{0.242224,0.643275,0.304449}
\definecolor{plots4}{rgb}{0.76444,0.444112,0.824298}
\definecolor{plots5}{rgb}{0.675544,0.555662,0.0942343}
\definecolor{plots6}{rgb}{0.00000482118,0.665759,0.680997}
\definecolor{plots7}{rgb}{0.930767,0.367477,0.57577}
\newcommand{\legendbox}[1]{%
\begin{tcolorbox}[nobeforeafter,left=1pt,right=1pt,top=2pt,bottom=2pt,boxsep=0mm,colback=#1!50!white,
                  colframe=#1,
                  width=18pt,
                  arc=3pt, auto outer arc,
                 ]
\end{tcolorbox}}

\graphicspath{{./}{./figures/}}
\newcommand{\mypara}{\paragraph}
\usepackage{enumitem}

\usenatbib
\usehyperref

\makeatletter
\def\blfootnote{\xdef\@thefnmark{}\@footnotetext}
\makeatother

\title{Low Rank Methods for Multiple Network Alignment}
\author{Huda Nassar, Georgios Kollias, Ananth Grama, David F.~Gleich}
\date{}

\begin{document}
\marginnote[500pt]{\fontsize{7}{9}\selectfont
Huda Nassar, Purdue University\\
\url{hnassar@purdue.edu}\\
\noindent Georgios Kollias, IBM\\
\url{gkollias@us.ibm.com}\\
\noindent Ananth Grama, Purdue University\\
\url{ayg@cs.purdue.edu}\\
\noindent David F.~Gleich, Purdue University\\
\url{dgleich@purdue.edu}
}


\maketitle

\begin{abstract} 
Multiple network alignment is the problem of identifying similar and related regions in a given set of networks. While there are a large number of effective techniques for pairwise problems with two networks that scale in terms of edges, these cannot be readily extended to align multiple networks as the computational complexity will tend to grow exponentially with the number of networks.In this paper we introduce a new multiple network alignment algorithm and framework that is effective at aligning thousands of networks with thousands of nodes. The key enabling technique of our algorithm is identifying an exact and easy to compute low-rank tensor structure inside of a principled heuristic procedure for pairwise network alignment called IsoRank. This can be combined with a new algorithm for $k$-dimensional matching problems on low-rank tensors to produce the alignment. We demonstrate results on synthetic and real-world problems that show our technique (i) is as good or better in terms of quality as existing methods, when they work on small problems, while running considerably faster and (ii) is able to scale to aligning a number of networks unreachable by current methods. We show in this paper that our method is the realistic choice for aligning multiple networks when no prior information is present. 
\end{abstract}

\section{Introduction}


%

Pairwise global network alignment (PNA) is the problem of matching pairs of nodes in two input graphs such that the pairing identifies common structures in both graphs. Algorithms for and applications of this problem are extensively discussed in the literature~\cite{Feizi-2016-spectral,Kuchaiev-2010-topological,Malod-Dognin-2015-lgraal,Patro-2012-ghost,Atias-2012-network-alignment,singh2008-isorank-multi,Bayati-2013-MessagePassing,Klau-2009-netalignmr,Nassar-2018-spectral,Langs-2010-functional-alignment}.
A more general problem is that of multiple global network alignment (MNA)~\cite{FUSE,Liao-IsorankN,Malmi-2017-flan}, where we are interested in finding a large subgraph present in more than two input networks. Applications of this routine arise in comparative proteomics (where the networks are protein interactions from multiple species), entity resolution (where the networks reflect different records), subject registration (where the networks reflect multiple measured views), and other applied machine learning tasks.

Both PNA and MNA are NP-hard generalizations of the subgraph isomorphism problem, and MNA is a harder problem in practice due to the combinatorial explosion of possible aligned pairs. As an illustration of this point, consider a common strategy in PNA algorithms~\cite{Klau-2009-netalignmr,Kollias-2012-NSD,Feizi-2016-spectral,Nassar-2018-spectral,Bayati-2013-MessagePassing,Patro-2012-ghost}: (i) score each potential matched pair of nodes between the graphs based on a topological similarity measure; and (ii) perform a maximum weight bipartite matching (or a closely related algorithm) on the set of scores. Simple extensions of these principled procedures to MNA with $k$ networks cannot easily scale to more than a handful of networks because the set of data in step  (i) becomes $O(n^k)$ when each network has $O(n)$ nodes, and (ii) the obvious generalization of max weight bipartite matching is $k$-dimensional matching, which is NP-complete for $k \ge 3$~\cite{Karp-1972}. As an alternative, there are approximation algorithms for $k$-dimensional matching~\cite{Kann-1991-3d-matching}.

Despite the computational difficulty, there are a few algorithms that navigate the computational and memory requirements. A straightforward solution is to consider sequences of pairwise network alignment problems, or to use pairwise network alignment data to infer multi-network alignments. Another straightforward solution is to restrict the set of possible alignments to those inferred through prior information or metadata about the nodes. Such information often speeds up the computation drastically and guides the algorithm to a meaningful solution~\cite{Malmi-2017-flan}. In this paper, we focus on the case when such information is not present and there is no reduction to pairwise data. To the best of our knowledge, ours is the first multiple network alignment algorithm that can scale to thousands of networks with thousands of nodes in a reasonable runtime (about 3 hours for 1000 networks with 1000 nodes). In this regime existing techniques take too long, run out of memory, or give bad results.

The two main technical innovations are (i) a specific multi-network generalization of the pairwise network alignment algorithm IsoRank~\cite{singh2008-isorank-multi} that enables us to compute a representation of the $O(n^k)$, $k$-way alignment data efficiently, and (ii) an extremely efficient $k$-dimensional matching algorithm with an a-posteriori approximation bound when the matching information is given by a low-rank  tensor. We summarize our findings and contributions here:
\begin{itemize}[leftmargin=*]
\item We generalize the IsoRank algorithm to multiple networks and show that the solution can be represented by a multidimensional tensor that can be explicitly written in terms of low-rank nonnegative factors that are easy to compute (Section~\ref{sec:multiisorank}).
\item We present a new $k$-dimensional matching algorithm for low-rank tensors with an a-posteriori approximation bound (Section~\ref{sec:lowrankmatch}).
\item We experimentally show that multiple network alignment is faster and higher-quality compared to performing multiple pairwise alignments when the number of networks grows (Section~\ref{sec:synthetic}).
\item We perform a case study on anonymized data from a collaboration network, where we show that aligning anonymized triplets of egonets can identify those triples with high Jaccard similarity, which can only be accurately computed from the de-anonymized data. (Section~\ref{sec:dblp}).
\end{itemize}


\section{Related work}
Existing MNA algorithms can be viewed in two classes. Biologically motivated algorithms are often designed to align protein-protein interaction networks, and topological algorithms are more generic and try to exploit the network structure. We review each of these classes briefly.

\mypara{Biological algorithms}
In biology, there is a need to discover new relationships between proteins, and MNA can be used as a tool to study these connections~\cite{singh2008-isorank-multi}. The networks to be aligned are often protein protein interaction networks (PPIs) of different species, and the idea is to use the alignment to learn new information about the less studied species. In these cases, there are several measures to compare the proteins independently of the network interaction structure, such as by evaluating the sequence similarity of their genetic codings. Biological algorithms are designed with this piece of information in mind, such as MultiMagna++~\cite{Vijayan-2017-Multimagna}, which uses a genetic algorithm that works directly with the multi-way alignment permutations and uses objective or fitness functions that utilize the biological information. 

IsoRank~\cite{singh2008-isorank-multi} and IsoRankN~\cite{Liao-IsorankN} were some of the earliest MNA algorithms. These computed pairwise topological similarity scores between each pair of networks and then assembled the result into a multiple alignment in a variety of ways. They can be related back to a complete $k$-partite network representation of all the pairwise alignment information. 
A more recent algorithm, FUSE~\cite{FUSE} uses protein sequence similarity to build the $k$-partite representation of the problem and then uses non-negative matrix trifactorization to incorporate network structure into the overall alignment.


\mypara{Topological algorithms}
There are two state of the art algorithms introduced in~\cite{Malmi-2017-flan}: \emph{FLAN} and \emph{PROGNATALIE++}. The FLAN method is based on generalizing the concept of the facility location problem and is a good way to utilize prior information about possible relationships (such as in entity resolution in their case). We compare against \emph{PROGNATALIE++} below, which extends the PNA algorithm Natalie proposed by Klau et al.~\cite{Klau-2009-netalignmr}.  \textit{PROGNATALIE++} proceeds by solving the multiple network alignment problem progressively, by aligning the first two networks, and then folding in the third network using the existing match, etc. This involves solving $k-1$ PNA problems.
\mypara{The need for new methods}
To run these algorithms on networks where no prior similarity measures is available, one can assume that all pairs of nodes are similar and assign them the same score. Such an approach empirically fails in producing meaningful results for the algorithms FUSE, IsoRankN, and MultiMagna++~\cite{FUSE,Liao-IsorankN,Vijayan-2017-Multimagna}. In contrast, PROGNATALIE++ and FLAN are more resistant to the absence of this information, but the running time of these algorithms is extreme as they are solving an NP-hard problem at each step.

\section{Multiple Network Alignment and An Exact Low-Rank Method}
\label{sec:multiisorank}
The multiple network alignment problem can be formulated for three undirected networks as: 
\begin{fullwidth}
\begin{equation}
 \MAXtwo{}{\sum_{i,j,k} \sum_{r,s,t} A_{ir} B_{js} C_{kt} \cX_{ijk} \cX_{rst}}{\sum_{j,k} \cX_{ijk} \le 1 \text{ for all $i$}; \sum_{i,k} \cX_{ijk} \le 1 \text{ for all $j$}; \sum_{i,j} \cX_{ijk} \le 1 \text{ for all $k$}}{\cX_{i,j,k} \in \{0, 1\} \text{ for all $i,j,k$}.}
\end{equation}
\end{fullwidth}
Here $\cX_{ijk}=1$ indicates that node $i$ in network $A$ matches to node $j$ in network $B$ and node $k$ of network $C$, $\mA, \mB, \mC$ are adjacency matrices for the three networks, and the number of vertices of these networks give the summation limits in the above expression. The objective function can be read as nodes $i, j, k$ are matched and we have edges $(i,r)$ in $A$, $(j,s)$ in $B$, and $(k,t)$ in $C$ which are all simultaneously preserved if we also match $r,s,t$. That is, the product of all of these expressions is 1 when all the edges exist and they match, and $0$ otherwise. The extension to $k$ networks will be straightforward once we introduce some notation. 

If we write $\vx = \tvec(\cmX)$, i.e $\vx$ is the large vector representation of the tensor data $\cmX$, then the objective function is: $\vx^T(\mC \kron \mB \kron \mA)\vx$. (This is an instance of the mixed-product property for Kronecker products and tensors, see, e.g. equation 12.4.19 in~\cite{Golub-2013-book}.) The constraints can be written in terms of the tensor \emph{flattening or unfolding} operator $\flat_j$ that turns $\cmX$ into a matrix by unfolding along dimension $j$ (see~\cite[Section 12.4.5]{Golub-2013-book}~and~\cite{Draisma-2014-bounded-rank}). Then, we have the three and $k$-network problems: 
\begin{fullwidth}
\[ 
  \!\!\!\MAXtwo{}{\vx^T (\mC \kron \mB \kron \mA) \vx}{ \flat_1(\cmX) \ve \le \ve; \flat_2(\cmX) \ve \le \ve; \flat_3(\cmX) \ve \le \ve}{\cX_{i,j,k} \in \{0, 1\} \text{ for all $i,j,k$.}}\!\!
  \MAXtwo{}{\vx^T (\mA_k \kron \cdots \kron \mA_1) \vx}{ \flat_1(\cmX) \ve \le \ve; \ldots ; \flat_k(\cmX) \ve \le \ve}{\cX_{i,j,\ldots,k} \in \{0, 1\} \text{ for all indices}.} 
\]
\end{fullwidth}
Here $\ve$ is the vector of all ones of appropriate dimension. Throughout, we frequently interchange between tensor representations of data $\cmX$ and their vectorized representations $\vx = \tvec(\cmX)$. 

Note that, if we were to relax to real-values and heuristically change the constraints to $\normof[2]{\vx} = 1$, then the solution is the eigenvector of $\mC \kron \mB \kron \mA$ with largest eigenvalue. This eigenvector could then be reshaped and input to a 3d matching routine to produce a multiple network alignment. In practice, this technique needs a number of improvements even for the pairwise case~\cite{Feizi-2016-spectral}, and these are non-trivial to adapt to the multiple network case, which is discussed further in the conclusion. Instead, we adapt the IsoRank methodology, and specifically, the network similarity decomposition (NSD) method~\cite{Kollias-2012-NSD} to compute IsoRank, which will easily scale to multiple networks; we explain these now.

Now we show that this formulation is closely related to optimizing over hyper-permutation matrices. Consider aligning three networks, the idea is to permute symmetric $\mC$ to match symmetric $\mB$ and $\mA$, and permute $\mB$ and $\mA$ to match $\mC$. This yields the objective:  
\begin{equation}
\min_{\cmP} \|  \cmP \times_1 \mC  - \cmP \times_2 \mB \times_3 \mA \|_F^2 = \min_{\cmP} \| (\mC \kron \mI \kron \mI)\vp - (\mI \kron \mB \kron \mA) \vp \|_2^2
\label{eq:hyperperm}
\end{equation}
where $\vp = \tvec(\cmP)$ and $\cmP$ is a hyper-permutation tensor and we use the $\tvec$ equivalences from ~\cite[Section 12.4.11]{Golub-2013-book}. This can be reworked into the objective $-2\vp^T (\mC \kron \mB \kron \mA) \vp + \vp^T(\mC \kron \mI \kron \mI)^2\vp + \vp^T (\mI \kron \mB \kron \mA)^2 \vp$. Since $\vp$ is a vectorized form of a permutation hyper-matrix, the two additive terms are almost a constant-expression if the networks have similar numbers of vertices and edges. Thus, we get an equivalence between the formulation in equation~\ref{eq:hyperperm} over hyper-matrices and ours if we neglect these terms.  This analysis extends to a variety of other ways to partition the set of networks into two groups. 

\paragraph{IsoRank} The IsoRank method for pairwise network alignment~\cite{singh2008-isorank-multi} used the PageRank vector $\vy$ of the graph with adjacency matrix $\mB \kron \mA$ (see~\cite{Gleich-2015-prbeyond} for more on this relationship) as a principled heuristic analogue of what we informally think of as a ``matching-biased eigenvector'' of $\mB \kron \mA$. Formally, let $\mD_A$ and $\mD_B$ be the diagonal degree matrices for graphs $A$ and $B$, then $\vy$ is given by the solution of the linear equations
\begin{equation}
 \vy = \alpha (\mB \mD_B^{-1} \kron \mA \mD_A^{-1}) \vy + (1-\alpha) \vh \quad \stackrel{\vy = \tvec(\mY)}{\Longleftrightarrow} \quad \mY = \alpha \mA \mD_A^{-1} \mY \mD_B^{-1} \mB + (1-\alpha) \mH.
\end{equation}
The value of $\alpha$ is typically chosen to be somewhere between $0.7$ and $0.9$ following~\cite{singh2008-isorank-multi} and the data $\vh$ or $\mH$ is either uniform (if there is no prior information about what might be a match) or chosen to represent some prior information. These equations can be solved without ever forming the Kronecker matrix, although, the data involved is still $O(n^2)$ for two $O(n)$ node graphs. Once we have the solution $\mY$, this can be turned into an alignment by solving a bipartite matching problem with $\mY$.

\paragraph{NSD} The NSD method specializes IsoRank in the case when $\mH$ is a low-rank matrix~\cite{Kollias-2012-NSD}, such as when we are using the uniform personalization term $\vh = \frac{1}{mn} \ve $, i.e., $\mH = \frac{1}{mn} \text{ones}(m,n)$ (where $A$ has $n$ vertices and $B$ has $m$ vertices and $\ve$ the vector of all ones of appropriate size). Thus, the relevant case for us is when $\mH$ is rank-1. Then there is an extremely efficient procedure to compute an exact low-rank representation of $\mY$. Suppose we initialize a fixed-point iteration for the PageRank linear system with $\mY\itn{0} = \mH = \vu \vv^T$ (because it is rank-1), and then $t^{\text{th}}$ iterate is given by:
\[
\mY^{(t)} = (1-\alpha) \sum_{i = 0}^{t-1} \alpha^i [(\mA \mD_A^{-1})^i \vu ] [(\mB \mD_B^{-1})^i \vv]^T + \alpha^{t} [(\mA \mD_A^{-1})^{t} \vu][(\mB \mD_B^{-1})^{t} \vv]^T.
\]
With some reorganization, this can be written: $\mY\itn{t+1} = \mU \mV^T$ for an $n$-by-$(t+1)$ matrix $\mU$ and an $m$-by-$(t+1)$ matrix $\mV$. The PageRank solution converges fast in the regime $\alpha \in [0.7, 0.9]$ and usually only $10$ iterations are enough. We now generalize this insight to multiple networks to handle multiple network alignment.

For multiple networks, the above formulation extends straightforwardly. We need to compute the PageRank vector on the network $\mA_k \kron \mA_{k-1} \kron \cdots \kron \mA_2 \kron \mA_1$. Since we have $k$ networks, the analogue of the matrix $\mY$ is now a $k$ dimensional tensor $\cmY$ that stores the PageRank measure between every possible combination $k$ of nodes coming from $k$ distinct networks. In words, we have $\cY(i_1,i_2,\hdots,i_k)$ denote the PageRank measure for the ``node'' representing an alignment between nodes $i_1$ from the first graph, $i_2$ from the second, $\hdots$, and node $i_k$ from the $k^{th}$ graph. Assume now that we have $k$ column stochastic adjacency matrices corresponding to $k$ networks. Call them $\mP_1 = \mA \mD_A^{-1}, \mP_2 = \mB \mD_B^{-1}, \hdots, \mP_k$. The massive PageRank vector we are interested in is given  by: 
\begin{equation}
 \vy = \alpha (\mP_k \kron \mP_{k-1} \kron \cdots \kron \mP_2 \kron \mP_1) \vy + (1-\alpha) \vh.
\end{equation}
Note that, although the problem $(\mP_k \kron \cdots \kron \mP_1)$ would never be formed, even creating the \emph{vector} $\vy$ would be impossible in terms of memory for all but the smallest problems as it takes $O(n^k)$ memory and there is no obvious sparsity to utilize. 
We study the case that $\vh = \vu_k \kron \vu_{k-1} \kron \cdots \kron \vu_1$, which corresponds to assuming that the tensor representation $\cmH$ would be rank 1. In this instance, we can proceed akin to the NSD scenario. We also start the iteration with $\vy^{(0)} = \vh =\vu_k \kron \vu_{k-1} \kron \cdots \kron \vu_1$.
Then, the first iterate is:
\[
\begin{aligned} 
\vy^{(1)} & = \alpha (\mP_k \kron \cdots \kron \mP_1) \vy^{(0)} + (1-\alpha) \vy^{(0)} \\
& = \alpha (\mP_k \vu_k \kron \hdots \kron \mP_1 \vu_1) + (1-\alpha) (\vu_k \kron \vu_{k-1} \kron \hdots \kron \vu_1)
\end{aligned}
\]
At step $t$, $\vy^{(t)}$ can be expressed as follows
\[
\vy^{(t)} = (1-\alpha) \sum_{i = 0}^{t-1} \alpha^i (\mP_k^i \vu_k \kron \hdots \kron \mP_1^i \vu_1) + \alpha^{t} \mP_k^{t} \vu_k \kron \hdots \kron \mP_1^{t} \vu_1
\]
Next, we can decompose the above equation. Form $k$ matrices $\mU_i$, such that
\begin{equation} \label{eq:Umats}
\mU_i = \bmat{c_0\mP_i^0 \vu_i & c_1 \mP_i^1 \vu_i & c_3 \mP_i^2 \vu_i & \hdots & c_{t-1} \mP_i^{t-1}\vu_i & c_t \mP_i^t \vu_i}
\end{equation}
where each $c_j$ is $((1-\alpha)\alpha^j)^{1/k}$ when $j \leq t-1$, and $c_t = \alpha^{t/k}$. Hence, $\vy^{t}$ can be rewritten as follows:
\[
\vy^{(t)} = \sum_{i = 0}^{t} \mU_k(:,i) \kron \mU_{k-1}(:,i) \kron \hdots \kron \mU_1(:,i) = \sum_{i = 0}^{t} \hat{\vy}^{(i)}
\]
where $\hat{\vy}^{(i)} = \mU_{k}(:,i) \kron \mU_{k-1}(:,i) \kron \hdots \kron \mU_1(:,i)$, and the notation $\mF(:,i)$ corresponds to the $i^{th}$ column of a matrix $\mF$. If we reshape into a tensor with $\tvec(\cmY_i) = \hat{\vy}^{(i)}$, then $\cmY\itn{t} = \sum_{i=0}^{t} \cmY_i$. We can thus deduce that $\cmY\itn{t}$ is a sum of $t+1$ rank-1 tensors. (Formally, the matrices $\mU_1, \ldots, \mU_k$ are the CP factors of $\cmY\itn{t}$~\cite[Section 12.5.4]{Golub-2013-book}.) What remains in our procedure is a way to turn this low-rank representation into an alignment by running a matching algorithm. 
\section{K-Dimensional Matching with Low-Rank Factors}
In this section, we discuss two approaches to solve the $k$-dimensional matching problem: 
\begin{equation}
 \MAXone{}{\sum_{i,j,\ldots,\ell} \cT(i,j,\ldots,\ell) \cX(i,j,\ldots,\ell)}{\flat_1(\cmX) \ve \le \ve; \ldots; \flat_k(\cmX) \ve \le \ve; \cX(i,j,\ldots,\ell) \in \{0, 1\}}
\end{equation}
when $\cmT$ is given by a non-negative rank-$t$ representation: 
\begin{equation} \label{eq:low-rank-T}
\begin{aligned}
& \textstyle \cT(i,j,\ldots,\ell) = \sum_{t=1}^r U_1(i,t) U_2(j,t) \cdots U_k(\ell,t)  \\
& \textstyle \qquad \Leftrightarrow   \cmT = \sum_{i=1}^t \cmT_i \text{ where } \tvec(\cmT_i) = \mU_k(:,i) \kron \mU_{k-1}(:,i) \kron \cdots \kron \mU_1(:,i)
\end{aligned}
\end{equation}
The first builds on an algorithm for low-rank bipartite matchings from~\cite{Nassar-2018-spectral}. The second builds on algorithms for progressive alignment~\cite{Malmi-2017-flan} and $k$-partite alignment problems~\cite{FUSE,He-2000-kpartite}.

\subsection{An a-posteriori approximation bound from the best single-rank alignment}
\label{sec:lowrankmatch}
We proceed to show a new $k$-dimensional matching algorithm that can be applied on tensors represented as low-rank factors. The idea is that we use each rank-1 factor $\cmT_i$ to generate a single $k$-dimensional matching. Then we provide an a-posteriori bound on the best alignment in this set. In practice, these bounds are very good and provide approximation factors around $1.08$ (see the supplementary material figure~\ref{fig:dblp_histogram}). 
The techniques extend~\cite{Nassar-2018-spectral} for the bipartite matching case. To do so, we first need a specific generalized rearrangement inequality for $k$ sequences. Generalized forms of the rearrangement inequality are often posed as a homework problem as their proof follows an extension to the proof by induction of the inequality for two sequences. For completeness, we provide a full proof in the supplementary material (section~\ref{sec:gen_rearrange}) and state the inequality here.

\mypara{Generalized Rearrangement Inequality.} Assume that we have $k$ sequences of numbers that are all positive. Let $x_i^{(j)}$ denote the $i^{th}$ element in the $j^{th}$ sequence and assume that $x_1^{(j)} \leq x_2^{(j)} \leq \hdots x_n^{(j)}$ for all sequences. The generalized rearrangement inequality guarantees that:
\[
\sum_{i=1}^n \prod_{j=1}^{k} x_i^{(j)} \geq \sum_{i=1}^n x_i^{(1)} \prod_{j=2}^k x_{\sigma_j(i)}^{(j)}
\]
where $\sigma_j$ is any permutation function corresponding to the $j^{th}$ sequence.

Now, assume that we have a \textit{k-dimensional} tensor $\cmT$ of the form~\eqref{eq:low-rank-T}. 
For each rank-1 tensor $\cmT_i$, the generalized rearrangement inequality guarantees the best matching on it can be computed by sorting the vectors $\mU_1(:,i), \ldots, \mU_k(:,i)$ in decreasing order and aligning the elements. (We find it helpful to think of the pairwise, matrix, case where $\mT_i = \vu \vv^T$ and the sorting is simple to see.) Let the binary-valued tensors $\cmM_i$ of size $n_1 \times n_2 \times \hdots n_k$ store the matching corresponding to $\cmT_i$ tensor, i.e., $\cM(j_1,j_2,\hdots,j_k) = 1$ if $(j_1,j_2,\hdots,j_k)$ is a match, and $0$ otherwise.  In the supplement, we prove the following: 

\textbf{Result.} \emph{Consider the best $k$-dimensional matching from the set $\cmM_1, \ldots, \cmM_t$, then this is a $D$-approximation to the best $k$-dimensional matching, where $D$ is an aposterori computable bound.}

\subsection{A progressive alignment}\label{sec:progmatch}

The bounds given by the low-rank matching algorithm (above) are often very good (around $1.08$, see the supplement). In practice we found the following procedure to give better results in terms of the overall multiple network alignment objective. The inspiration for this algorithm is the progressive nature of both \emph{ProgNatalie++} and \emph{FUSE}~\cite{Malmi-2017-flan,FUSE}, and a progressive algorithm for the $k$-partite matching problem~\cite{He-2000-kpartite}. For three networks (a three-mode tensor), the idea is: align (via bipartite matching) the first two modes (networks). Then, use the alignment between the first two modes to produce a new bipartite alignment problem to fold in the third mode. That is, if we know that node $i_1$ in network 1 matches to $i_2$ in network $2$, then we can look at the entries $\cmT(i_1,i_2,:)$ to determine the best match for $(i_1,i_2)$ in the third network.  These entries also have low-rank structure. This can be done via $k$ bipartite matching calls in our low-rank framework, and it is easiest to state the overall procedure as an algorithm. We briefly studied optimizing the ordering of alignment, but this did not seem to yield large differences. 

\begin{figure}[h]
\begin{lstlisting}[escapeinside={(*}{*)}]
(*\textbf{Input}*): (*$\mU_1, \mU_2, \hdots, \mU_k$*); (*\textbf{Output}*): Matching M with k columns and matches in rows
a,b = bipartitematching((*$\mU_1^{} \mU_2^T$*)) % match first two modes and return matches a and b.
M[:,1:2] = [a,b] 
for i = 3 to k # 
  % generate the matching information with $i-1$ modes matched
  U = U$_1$[M[:,1],:] $\odot$ U$_2$[M[:,2],:] $\odot \cdots \odot$ U$_{i-1}$[M[:,i],:] 
  % using element-wise/Hadamard product $\odot$
  a,b = bipartitematching((*$\mU^{} \mU_i^T$*)) % match in the ith mode
  M = M[a,:]; M[:,i] = b % permute and extend the matching 
end
\end{lstlisting}
\vspace{-1em}
\caption{Pseudocode for the progressive $k$-dimensional matching algorithm}
\label{fig:alg_kpartite} 
\end{figure}


\section{Experiments}
To evaluate our proposed algorithm, we perform a series of experiments (i) on synthetically generated networks where we can easily vary parameters to understand how the algorithms behave, (ii) on the problem of aligning snapshots of a temporally evolving network of internet routers, and (iii) on inferring high triangle Jaccard similarity in anonymized egonets. 

We precisely state the parameters of the various methods we consider here, including some obvious baseline measures. We consider a few additional methods in the supplementary material including random alignments (section~\ref{sec:additional_algs}). We also tried two software packages \emph{IsoRankN} and \emph{FUSE} for these problems. These methods all returned empty alignments, which we believe is due to our lack of \emph{prior} or \emph{biological} information to guide the method. 

\mypara{Pairwise.} A simple way to align multiple networks is to run a pairwise network alignment for all pairs of networks and extract any consistent alignment. For instance, if the following three pairs appeared while aligning the three networks $\mG_A, \mG_B, \mG_C$, $(a_1, b_3), (b_3,c_9), (a_1,c_9)$, we treat the triplet $(a_1,b_3,c_9)$ as a match. For choosing the right pairwise method to employ in this paradigm, we wanted a pairwise method that does not rely on prior similarity scores, thus we chose the recent low rank spectral network alignment by~\cite{Nassar-2018-spectral}. 

\mypara{By degree.} This method is intuitive since we would expect that high degree nodes match to each other. For each network, sort the nodes according to their degrees, and then match the top degree nodes with each other until no more nodes are left in one of the networks.


\mypara{MultiLR-D} This is our algorithm where we compute the matrices $\mU_i$ from~\eqref{eq:Umats} with 8 iterations and $\alpha=0.8$ then the final alignment is extracted by our $D$-approximation (Section~\ref{sec:lowrankmatch}). See the supplementary information for a study on why 8 iterations is enough (section~\ref{sec:eight_iters}).

\mypara{MultiLR-Prog} This is our algorithm where the $\mU_i$ are from~\eqref{eq:Umats} with 8 iterations  and $\alpha=0.8$ and the final match is determined by the progressive method (Figure~\ref{fig:alg_kpartite}). The bipartite matching problems are themselves solved via a low-rank bipartite matching procedure from~\cite{Nassar-2018-spectral} (with parameter $b=10$). 

\mypara{MultiLR-Prog+} This is the same as MultiLR-Prog, but where we replace the element-wise multiplication from Figure~\ref{fig:alg_kpartite} (line 6) with a mixture model for $\mU$. Specifically we use $\mU = (1/2) \mU^*/\text{sum}(\mU^*) + (1/2) \mU^+/\text{sum}(\mU^+)$ where  $\mU^*$ is the matrix computed on line $6$ and $\mU^+$ is the matrix computed on line 6 with element-wise multiplication replaced with element-wise addition. Empirically (and by accident), we found that this strategy performed more consistently with large numbers of networks; theoretically, it is more akin to treating the alignment data as finding a combination of $k$-dimensional matches and dense $k$-partite regions as in~\cite{FUSE,Liao-IsorankN}.

\mypara{ProgNatalie++ and ProgNatalie++ with prior} We use \emph{ProgNatalie++} from~\cite{Malmi-2017-flan} using a uniform prior for small problems. This does not scale with a reasonable runtime (we ran problems with 100 nodes and 5 networks for a day without completing), and so we also consider using the union of alignments produced by our low-rank factors (Section~\ref{sec:lowrankmatch}) as the prior. In this case, the algorithms complete in a reasonable amount of time (an hour for 5 networks with 100 nodes) because of the constrained matching space.

We use a few evaluation metrics to discuss the resulting alignments. When there is a true alignment known among the set of networks, then we compute \emph{\textbf{degree weighted recovery}}, which is the number of correct pairs, scaled by the degrees of the nodes in the network. We often found that the algorithms would align large portions of the network well, but make mistakes on regions of ambiguous degree-1 nodes (or other automorphic regions of the graphs). Consequently, this measure places more emphasis on high degree regions. The pairwise nature also protects against a single mistake in, say, 100 networks ruining the other 99 correctly aligned results. The formal measure involves some ancillary notation. Let $D_j$ be the sum of all degrees in network $j$. The weight of a pair of vertices in network $j$ and $k$ in a pair of networks is $w(v_j, v_k) = (\text{degree}(v_j)+\text{degree}(v_k))/(D_j+D_k)$; the expression $\text{correct}(v_j,v_k)$ is one if node $v$ from network $j$ should be aligned to node $v$ from network $k$; the \emph{score} of a single alignment of vertices between all networks is: 
\[
\textstyle \text{score}(v_1,v_2,\hdots,v_k) = \binom{k}{2}^{-1} ( \sum_{j=1}^{k}\sum_{h=j+1}^{k} w(v_j,v_h)\text{correct}(v_j,v_h) )
\]
The overall degree weighted recovery score is simply the sum scores for each alignment set. (These scores are scaled to sum to 1 for a perfect alignment of isomorphic networks.) 

The \emph{\textbf{normalized overlap}} of a set of networks $A_1, \ldots, A_k$ is the number of edges in the conserved region after alignment scaled by the number of edges of the largest graph. (Again, normalized overlap scores are between 0 and 1.). If $\tilde{\mA}_1, \ldots \tilde{\mA}_k$ are the adjacency matrices permuted via the alignment, then this is $\text{nnz}(\tilde{\mA}_1 \odot \cdots \odot \tilde{\mA}_k)/\text{max}(\text{nnz}(\mA_i), \ldots, \text{nnz}(\mA_k))$ where $\odot$ is the element-wise product.

\subsection{Erd\H{o}s-R\'enyi and preferential attachment graphs}
\label{sec:synthetic} 
In this first experiment, our goal is to study how well our algorithm recover solutions in a planted problem as we add more noise and how this changes as we vary the number of networks to be aligned. We consider \erdosrenyi and preferential attachment graphs with average degree $8$ (more details in the supplementary information section~\ref{sec:graph_construction}) as reference graphs, and then randomly delete edges to generate $k$ instances of the networks to align. In this case, the ground-truth alignment is known. 

\begin{fullwidthfigure}
\parbox{0.5\linewidth}{\centering \itshape As edge deletion varies \ldots }%
\parbox{0.5\linewidth}{\centering \itshape As we add networks \ldots }
\parbox[c]{0.25\linewidth}{\centering \footnotesize ~~~~Erd\H{o}s-R\'enyi}%
\parbox[c]{0.25\linewidth}{\centering \footnotesize ~~PA}%
\parbox[c]{0.25\linewidth}{\centering \footnotesize ~~~~Erd\H{o}s-R\'enyi}%
\parbox[c]{0.25\linewidth}{\centering \footnotesize ~~PA}
\includegraphics[width=0.25\linewidth]{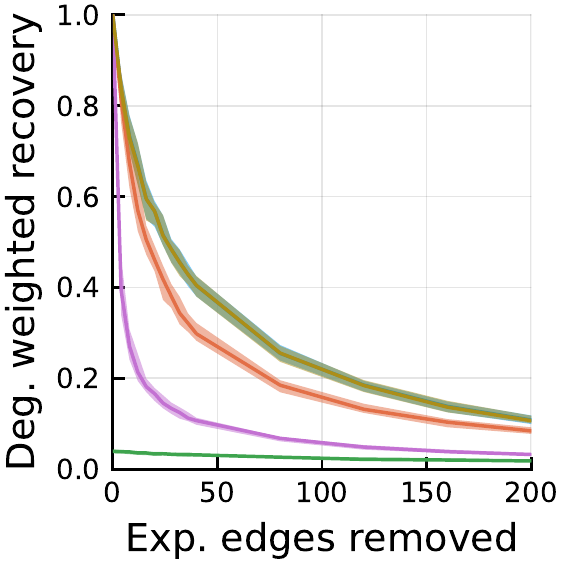}%
\includegraphics[width=0.25\linewidth]{PA/PA_PE_n5}%
\includegraphics[width=0.25\linewidth]{ER/ER_k}%
\includegraphics[width=0.25\linewidth]{PA/PA_k}

\parbox{\linewidth}{\centering \footnotesize
\legendbox{plots1}
MultiLR-Prog+ $\quad$ 
\legendbox{plots2}
Pairwise $\quad$ 
\legendbox{plots3}
Degree $\quad$ 
\legendbox{plots4}
MultiLR-D $\quad$ 
\legendbox{plots5}
MultiLR-Prog $\quad$ 
}
	\caption{(First two panels) As we increase the expected number of edges removed while aligning 5 networks, all methods recover fewer true matches and MultiLR-Prog and MultiLR-Prog+ are consistently the best where MultiLR-D does not do well. Note that MultiLR-Prog and MultiLR-Prog+ are overlapping here. (Last two panels) As we vary the number of networks to be aligned, all methods decay in quality except for MultiLR-Prog+ and MultiLR-D, with MultiLR-Prog+ consistently achieving the best result (the two right most figures). In all figures, the shaded areas represent the $20^{th}$ and the $80^{th}$ percentiles with these experiments run for 50 trials.}
	\label{fig:er_and_pa}
\end{fullwidthfigure}

For our first experiment, we consider using 5 networks with 500 nodes and vary the edge-deletion probability. The results from our methods and the baselines are shown in Figure~\ref{fig:er_and_pa} (first two panels). 
In both types of graphs, both of our progressive low-rank methods achieved the best results, whereas MultiLR-D did not perform well as more edges were deleted. Although this method is the least expensive (see runtime discussion in the supplementary information section~\ref{sec:runtime}) and provides a theoretically strong bound on the matches (here, the highest value of D was 1.07) the method relies on a sorting procedure, which may mislead the matching when there are many numbers close to each other. 

For the second experiment, we also consider $500$ node networks again and consider aligning a growing number of networks with a fixed edge deletion probability $0.5/n$. This corresponds to the case where we expect good accuracy. The results are shown in  Figure~\ref{fig:er_and_pa} (last two panels) and show that MultiLR-Prog+ and MultiLR-D are the only methods that not sensitive to the number of networks. Because of this, MultiLR-D becomes a competitive method for large numbers of networks.

\subsection{Aligning real-world graph snapshots}
A representative use of our methods would be to align a set of snapshots of a real-world graph over time. 
Here we consider a dataset from~\cite{Leskovec-graphovertime-routers} which consists of snapshots of an Internet routers network at 733 time points. We consider two problems: aligning 5 random snapshots with the 25 highest degree nodes (where we are able to run existing methods) and aligning 5 random snapshots with the 100 highest degree nodes (where we can still run Prognatalie++ with our low-rank generated prior). In Figure~\ref{fig:routers}, we show a violin plot of the distribution of our results in terms of overlap and degree weighted recovery over 50 trials of 5 random snapshots. For the small run, we get comparable results to Prognatalie++, while running in less than 2 seconds vs. 40 minutes (see more timing in the supplementary info section~\ref{sec:runtime}). For the larger run, MultiLR-Prog and MultiLR-Prog+ achieve results that consistently outperform the pairwise baseline in terms of overlap.

\begin{fullwidthfigure}
			\begin{minipage}{0.5\linewidth}
				\centering \includegraphics[width=\linewidth]{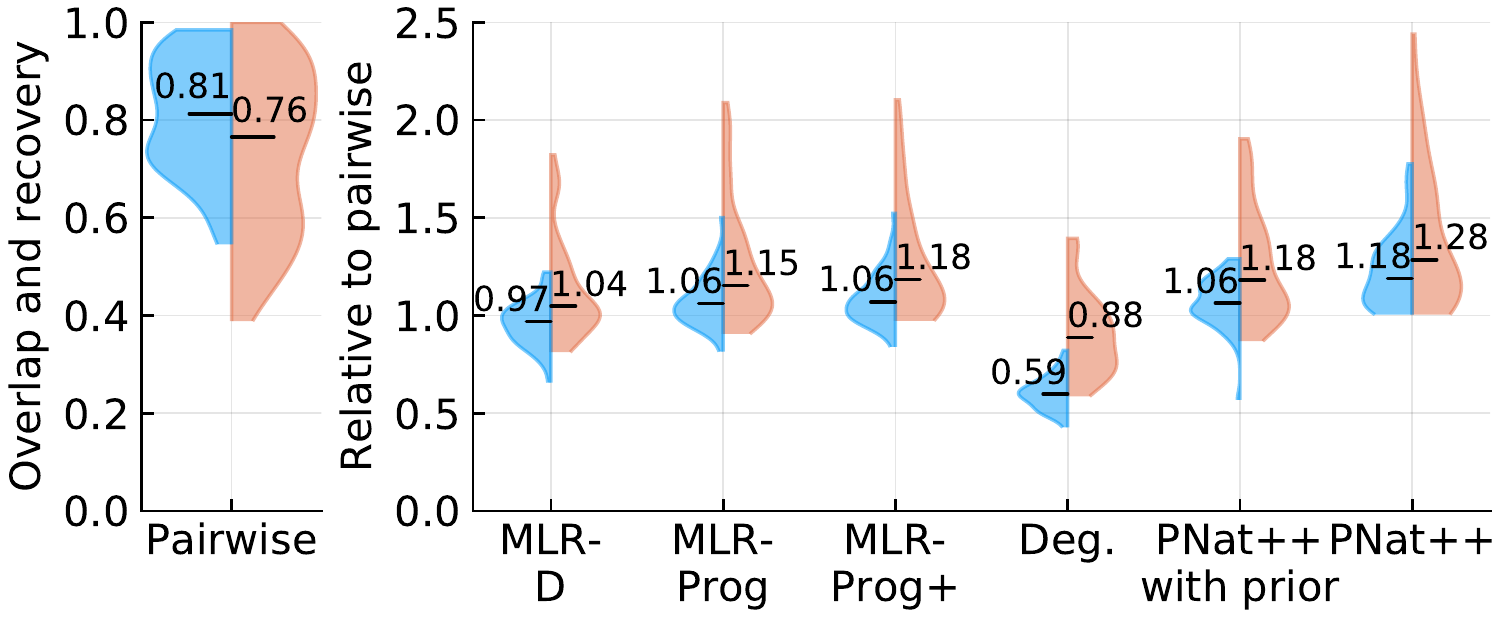}
				\footnotesize  \emph{Top 25 nodes}
			\end{minipage}%
			\begin{minipage}{0.5\linewidth}
				\centering \includegraphics[width=\linewidth]{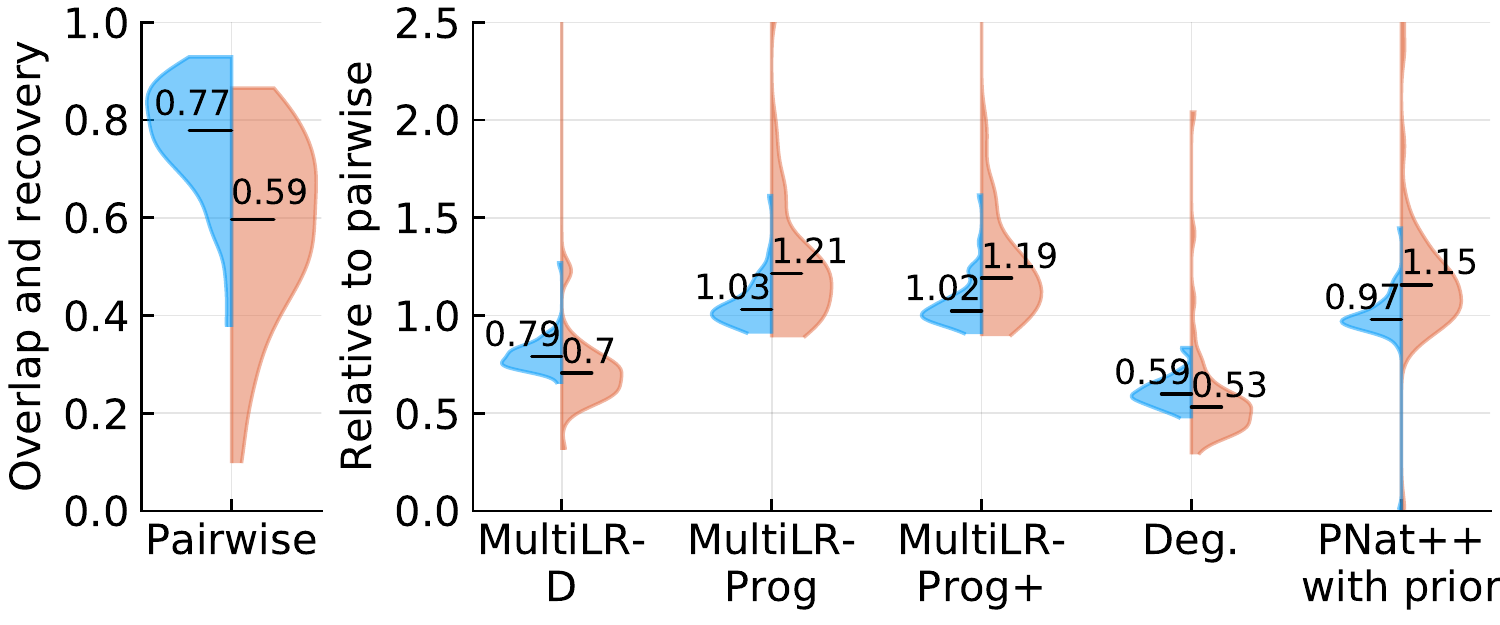}
				\footnotesize  \emph{Top 100 nodes}
			\end{minipage}
			
			\caption{We consider 50 trials of aligning 5 real-world router graphs and show a violin plot (with the median flagged) of the \textcolor{plots1}{degree weighted recovery (blue)} and \textcolor{plots2}{normalized overlap (red)} side by side for the pairwise method. For the other methods, we show values relative to the pairwise scores. These results show that we are almost as good as the existing state of the art method ProgNatalie++ on the small problems, whereas our methods run faster, and we can scale to larger problems.}
			\label{fig:routers}
\end{fullwidthfigure}

\subsection{Aligning anonymized egonets}
\label{sec:dblp}
Next, we use our multiple network alignment algorithm to align anonymized egonets of the collaboration network DBLP~\cite{Esfandiar-2010-Katz}. This experiment is inspired by one in~\cite{Nassar-2018-spectral}. In DBLP, the nodes are authors, and edges represent coauthorship. We consider whether or not multiple alignment could infer whether a group of three mutual coauthors (i.e. a triangle in the network) has high Jaccard similarity when we only know anonymized egonets from the original network. We use $\text{Jaccard}(a,b,c) = \frac{N(a) \cap N(b) \cap N(c)}{N(a) \cup N(b) \cup N(c)}$ where $N(a)$ is the set of neighbors of node $a$. For each triple of three coauthors with at least 100 other co-authors, we align the egonets using the MultiLR-D method and measure the normalized overlap. The results in Figure~\ref{fig:dblp} show that we can easily infer high-Jaccard similarity whereas pairwise techniques cannot. This experiment entails aligning 425,388 triplets of networks and MultiLR-D runs in about 1.5 hours whereas the pairwise method takes a little over 4 hours to finish. To ensure that we could be confident that high-overlap implies high-Jaccard, we show that random triples are unlikely to have high normalized overlap in the final figure panel.

\begin{tuftefigure}
\centering
	\begin{minipage}{0.33\linewidth}
		\centering \includegraphics[width=\linewidth]{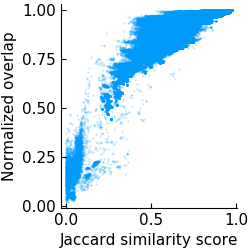}\\
		\footnotesize \emph{(a) Using MNA}
	\end{minipage}%
	\begin{minipage}{0.33\linewidth}
		\centering \includegraphics[width=\linewidth]{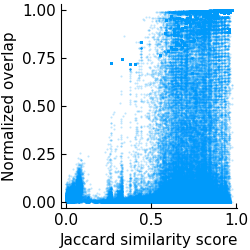}\\
		\footnotesize \emph{(b) Using pairwise}
	\end{minipage}%
	\begin{minipage}{0.33\linewidth}
		\centering \includegraphics[width=\linewidth]{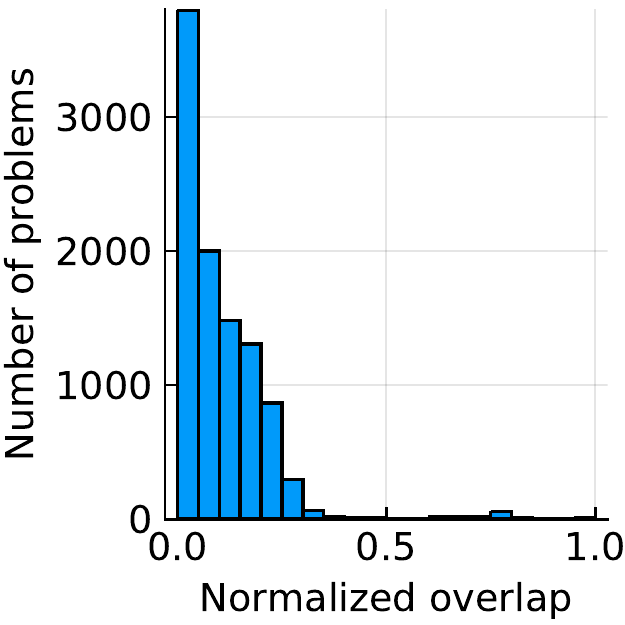}\\
		\footnotesize \emph{(c) MNA on random networks}
	\end{minipage}
	\vspace{-0.5\baselineskip}
	\caption{Figures a and b show the normalized overlap of the aligned three egonets using multiple network alignment and pairwise respectively. These two figures show that when using multiple network alignment, normalized overlap track the Jaccard similarity scores whereas the pairwise method fails to show that. Figure c shows that the opposite is true as well. For a random set of three networks, the normalized overlap is is less than 0.25 in the majority of experiments.}
	\label{fig:dblp}
\end{tuftefigure}

\section{Discussion and future work}
Having a method that accurately and scalably aligns large numbers of networks opens a number of new dimensions in applied machine learning. In ongoing work, we are studying how to use this in terms of aligning graphs derived from functional MRI data. In terms of the current method, we wish to better understanding for why MultiLR-Prog+ outperformed MultiLR-Prog. Our working hypothesis is that the element-wise addition (compared with multiplication) gives the method resilience to mistakes made early in the progressive process. More broadly, the EigenAlign framework~\cite{Feizi-2016-spectral} is superior to the IsoRank framework for pairwise alignment. The ideas here apply to a multi-network generalization of EigenAlign, however, the analogous tensor $\cmY$ would have a Tucker-style factorization instead of the CP-factorization we get for MultiLR. Crucially, the Tucker factorization needs a $t^k$-element core that would limit scalability to small $k$, and we need new $k$-dimensional matching methods for these. 

\section*{Acknowledgements} 
The authors were supported by NSF CCF-1149756, IIS-1422918, IIS-1546488, CCF-0939370, DARPA SIMPLEX, and the Sloan Foundation.

\bibcolumns=3
\begin{fullwidth} 
\bibliographystyle{dgleich-bib}
\bibliography{s999-refs}

\begin{thebibliography}{23}
\providecommand{\natexlab}[1]{#1}
\providecommand{\bibextraformatting}{\relax}
\bibextraformatting

\bibitem[\protect\citeauthoryear{Atias and
  Sharan}{2012}]{Atias-2012-network-alignment}
N.~\textsc{Atias} and R.~\textsc{Sharan}.
\newblock \href{http://dx.doi.org/10.1145/2160718.2160738}{\emph{Comparative
  analysis of protein networks: hard problems, practical solutions}}.
\newblock Commun. ACM, 55~(5), pp. 88--97, 2012.
\newblock \href {http://dx.doi.org/10.1145/2160718.2160738}
  {\normalcolor\path{doi:10.1145/2160718.2160738}}.

\bibitem[\protect\citeauthoryear{Bayati
  et~al.}{2013}]{Bayati-2013-MessagePassing}
M.~\textsc{Bayati}, D.~F. \textsc{Gleich}, A.~\textsc{Saberi}, and
  Y.~\textsc{Wang}.
\newblock
  \href{http://dx.doi.org/10.1145/2435209.2435212}{\emph{Message-passing
  algorithms for sparse network alignment}}.
\newblock ACM Trans. Knowl. Discov. Data, 7~(1), pp. 3:1--3:31, 2013.
\newblock \href {http://dx.doi.org/10.1145/2435209.2435212}
  {\normalcolor\path{doi:10.1145/2435209.2435212}}.

\bibitem[\protect\citeauthoryear{Draisma and
  Kuttler}{2014}]{Draisma-2014-bounded-rank}
J.~\textsc{Draisma} and J.~\textsc{Kuttler}.
\newblock \href{http://dx.doi.org/10.1215/00127094-2405170}{\emph{Bounded-rank
  tensors are defined in bounded degree}}.
\newblock Duke Mathematical Journal, 163~(1), pp. 35--63, 2014.
\newblock \href {http://dx.doi.org/10.1215/00127094-2405170}
  {\normalcolor\path{doi:10.1215/00127094-2405170}}.

\bibitem[\protect\citeauthoryear{Esfandiar et~al.}{2010}]{Esfandiar-2010-Katz}
P.~\textsc{Esfandiar}, F.~\textsc{Bonchi}, D.~F. \textsc{Gleich},
  C.~\textsc{Greif}, L.~V.~S. \textsc{Lakshmanan}, and B.-W. \textsc{On}.
\newblock \href{http://dx.doi.org/10.1007/978-3-642-18009-5_13}{\emph{Fast Katz
  and Commuters: Efficient Estimation of Social Relatedness in Large
  Networks}}, pp. 132--145.
\newblock Springer Berlin Heidelberg, Berlin, Heidelberg, 2010.
\newblock \href {http://dx.doi.org/10.1007/978-3-642-18009-5_13}
  {\normalcolor\path{doi:10.1007/978-3-642-18009-5_13}}.

\bibitem[\protect\citeauthoryear{Feizi et~al.}{2016}]{Feizi-2016-spectral}
S.~\textsc{Feizi}, G.~\textsc{Quon}, M.~R. \textsc{Mendoza},
  M.~\textsc{M{\'{e}}dard}, M.~\textsc{Kellis}, and A.~\textsc{Jadbabaie}.
\newblock \href{http://arxiv.org/abs/1602.04181}{\emph{Spectral alignment of
  networks}}.
\newblock arXiv, cs.DS, p. 1602.04181, 2016.

\bibitem[\protect\citeauthoryear{Gleich}{2015}]{Gleich-2015-prbeyond}
D.~F. \textsc{Gleich}.
\newblock \href{http://dx.doi.org/10.1137/140976649}{\emph{{PageRank} beyond
  the web}}.
\newblock SIAM Review, 57~(3), pp. 321--363, 2015.
\newblock \href {http://dx.doi.org/10.1137/140976649}
  {\normalcolor\path{doi:10.1137/140976649}}.

\bibitem[\protect\citeauthoryear{Gligorijevic et~al.}{2016}]{FUSE}
V.~\textsc{Gligorijevic}, N.~\textsc{Malod-Dognin}, and N.~\textsc{Przulj}.
\newblock \href{http://dx.doi.org/10.1093/bioinformatics/btv731}{\emph{Fuse:
  multiple network alignment via data fusion}}.
\newblock Bioinformatics, 32~(8), pp. 1195--1203, 2016.
\newblock \href
  {http://arxiv.org/abs//oup/backfile/content_public/journal/bioinformatics/32/8/10.1093_bioinformatics_btv731/3/btv731.pdf}
  {\normalcolor\path{arXiv:/oup/backfile/content_public/journal/bioinformatics/32/8/10.1093_bioinformatics_btv731/3/btv731.pdf}},
  \href {http://dx.doi.org/10.1093/bioinformatics/btv731}
  {\normalcolor\path{doi:10.1093/bioinformatics/btv731}}.

\bibitem[\protect\citeauthoryear{Golub and van Loan}{2013}]{Golub-2013-book}
G.~H. \textsc{Golub} and C.~\textsc{van Loan}.
\newblock \emph{Matrix Computations}, Johns Hopkins University Press, 4th
  edition, 2013.

\bibitem[\protect\citeauthoryear{{He} et~al.}{2000}]{He-2000-kpartite}
G.~\textsc{{He}}, J.~\textsc{{Liu}}, and C.~\textsc{{Zhao}}.
\newblock \href{http://dx.doi.org/10.7155/jgaa.00021}{\emph{Approximation
  algorithms for some graph partitioning problems}}.
\newblock Journal of Graph Algorithms and Applications, 4~(2), pp. 1--11, 2000.
\newblock \href {http://dx.doi.org/10.7155/jgaa.00021}
  {\normalcolor\path{doi:10.7155/jgaa.00021}}.

\bibitem[\protect\citeauthoryear{Kann}{1991}]{Kann-1991-3d-matching}
V.~\textsc{Kann}.
\newblock \href{http://dx.doi.org/10.1016/0020-0190(91)90246-E}{\emph{Maximum
  bounded 3-dimensional matching is max snp-complete}}.
\newblock Information Processing Letters, 37~(1), pp. 27 -- 35, 1991.
\newblock \href {http://dx.doi.org/10.1016/0020-0190(91)90246-E}
  {\normalcolor\path{doi:10.1016/0020-0190(91)90246-E}}.

\bibitem[\protect\citeauthoryear{Karp}{1972}]{Karp-1972}
R.~M. \textsc{Karp}.
\newblock
  \href{http://www.cs.berkeley.edu/~luca/cs172/karp.pdf}{\emph{Reducibility
  among combinatorial problems}}.
\newblock In \emph{Proceedings of a symposium on the Complexity of Computer
  Computations, held March 20-22, 1972, at the {IBM} Thomas J. Watson Research
  Center, Yorktown Heights, New York.}, pp. 85--103. 1972.

\bibitem[\protect\citeauthoryear{Klau}{2009}]{Klau-2009-netalignmr}
G.~W. \textsc{Klau}.
\newblock \emph{A new graph-based method for pairwise global network
  alignment}.
\newblock BMC Bioinformatics, 10~(1), p. S59, 2009.

\bibitem[\protect\citeauthoryear{Kollias et~al.}{2012}]{Kollias-2012-NSD}
G.~\textsc{Kollias}, S.~\textsc{Mohammadi}, and A.~\textsc{Grama}.
\newblock \href{http://dx.doi.org/10.1109/TKDE.2011.174}{\emph{Network
  similarity decomposition (nsd): A fast and scalable approach to network
  alignment}}.
\newblock IEEE Trans. on Knowl. and Data Eng., 24~(12), pp. 2232--2243, 2012.
\newblock \href {http://dx.doi.org/10.1109/TKDE.2011.174}
  {\normalcolor\path{doi:10.1109/TKDE.2011.174}}.

\bibitem[\protect\citeauthoryear{Kuchaiev
  et~al.}{2010}]{Kuchaiev-2010-topological}
O.~\textsc{Kuchaiev}, T.~\textsc{Milenkovi{\'c}}, V.~\textsc{Memi{\v
  s}evi{\'c}}, W.~\textsc{Hayes}, and N.~\textsc{Pr{\v z}ulj}.
\newblock \href{http://dx.doi.org/10.1098/rsif.2010.0063}{\emph{Topological
  network alignment uncovers biological function and phylogeny}}.
\newblock Journal of The Royal Society Interface, 2010.
\newblock \href {http://dx.doi.org/10.1098/rsif.2010.0063}
  {\normalcolor\path{doi:10.1098/rsif.2010.0063}}.

\bibitem[\protect\citeauthoryear{Langs
  et~al.}{2010}]{Langs-2010-functional-alignment}
G.~\textsc{Langs}, P.~\textsc{Golland}, Y.~\textsc{Tie}, L.~\textsc{Rigolo},
  and A.~\textsc{Golby}.
\newblock
  \href{http://books.nips.cc/papers/files/nips23/NIPS2010_0078.pdf}{\emph{Functional
  geometry alignment and localization of brain areas}}.
\newblock In \emph{Proceedings of the 24th Annual Conference on Neural
  Information Processing Systems 2010; 1:1225-1233.}, pp. 1225--1233. 2010.

\bibitem[\protect\citeauthoryear{Leskovec
  et~al.}{2005}]{Leskovec-graphovertime-routers}
J.~\textsc{Leskovec}, J.~\textsc{Kleinberg}, and C.~\textsc{Faloutsos}.
\newblock \href{http://dx.doi.org/10.1145/1081870.1081893}{\emph{Graphs over
  time: Densification laws, shrinking diameters and possible explanations}}.
\newblock In \emph{Proceedings of the Eleventh ACM SIGKDD International
  Conference on Knowledge Discovery in Data Mining}, pp. 177--187. 2005.
\newblock \href {http://dx.doi.org/10.1145/1081870.1081893}
  {\normalcolor\path{doi:10.1145/1081870.1081893}}.

\bibitem[\protect\citeauthoryear{Liao et~al.}{2009}]{Liao-IsorankN}
C.-S. \textsc{Liao}, K.~\textsc{Lu}, M.~\textsc{Baym}, R.~\textsc{Singh}, and
  B.~\textsc{Berger}.
\newblock
  \href{http://dx.doi.org/10.1093/bioinformatics/btp203}{\emph{Isorankn:
  spectral methods for global alignment of multiple protein networks}}.
\newblock Bioinformatics, 25~(12), pp. i253--i258, 2009.
\newblock \href {http://dx.doi.org/10.1093/bioinformatics/btp203}
  {\normalcolor\path{doi:10.1093/bioinformatics/btp203}}.

\bibitem[\protect\citeauthoryear{Malmi et~al.}{2017}]{Malmi-2017-flan}
E.~\textsc{Malmi}, S.~\textsc{Chawla}, and A.~\textsc{Gionis}.
\newblock \href{http://dx.doi.org/10.1007/s10618-017-0505-2}{\emph{Lagrangian
  relaxations for multiple network alignment}}.
\newblock Data Min. Knowl. Discov., 31~(5), pp. 1331--1358, 2017.
\newblock \href {http://dx.doi.org/10.1007/s10618-017-0505-2}
  {\normalcolor\path{doi:10.1007/s10618-017-0505-2}}.

\bibitem[\protect\citeauthoryear{Malod-Dognin and
  Pr\v{z}ulj}{2015}]{Malod-Dognin-2015-lgraal}
N.~\textsc{Malod-Dognin} and N.~\textsc{Pr\v{z}ulj}.
\newblock \href{http://dx.doi.org/10.1093/bioinformatics/btv130}{\emph{L-graal:
  Lagrangian graphlet-based network aligner}}.
\newblock Bioinformatics, 31~(13), pp. 2182--2189, 2015.
\newblock \href
  {http://arxiv.org/abs/http://bioinformatics.oxfordjournals.org/content/31/13/2182.full.pdf+html}
  {\normalcolor\path{arXiv:http://bioinformatics.oxfordjournals.org/content/31/13/2182.full.pdf+html}},
  \href {http://dx.doi.org/10.1093/bioinformatics/btv130}
  {\normalcolor\path{doi:10.1093/bioinformatics/btv130}}.

\bibitem[\protect\citeauthoryear{Nassar et~al.}{2018}]{Nassar-2018-spectral}
H.~\textsc{Nassar}, N.~\textsc{Veldt}, S.~\textsc{Mohammadi},
  A.~\textsc{Grama}, and D.~F. \textsc{Gleich}.
\newblock \href{http://dx.doi.org/10.1145/3178876.3186128}{\emph{Low Rank
  Spectral Network Alignment}}, pp. 619--628.
\newblock 2018.
\newblock \href {http://dx.doi.org/10.1145/3178876.3186128}
  {\normalcolor\path{doi:10.1145/3178876.3186128}}.

\bibitem[\protect\citeauthoryear{Patro and Kingsford}{2012}]{Patro-2012-ghost}
R.~\textsc{Patro} and C.~\textsc{Kingsford}.
\newblock \href{http://dx.doi.org/10.1093/bioinformatics/bts592}{\emph{Global
  network alignment using multiscale spectral signatures}}.
\newblock Bioinformatics, 28~(23), pp. 3105--3114, 2012.
\newblock \href
  {http://arxiv.org/abs/http://bioinformatics.oxfordjournals.org/content/28/23/3105.full.pdf+html}
  {\normalcolor\path{arXiv:http://bioinformatics.oxfordjournals.org/content/28/23/3105.full.pdf+html}},
  \href {http://dx.doi.org/10.1093/bioinformatics/bts592}
  {\normalcolor\path{doi:10.1093/bioinformatics/bts592}}.

\bibitem[\protect\citeauthoryear{Singh et~al.}{2008}]{singh2008-isorank-multi}
R.~\textsc{Singh}, J.~\textsc{Xu}, and B.~\textsc{Berger}.
\newblock \href{http://dx.doi.org/10.1073/pnas.0806627105}{\emph{Global
  alignment of multiple protein interaction networks with application to
  functional orthology detection}}.
\newblock PNAS, 105~(35), pp. 12763--12768, 2008.
\newblock \href {http://dx.doi.org/10.1073/pnas.0806627105}
  {\normalcolor\path{doi:10.1073/pnas.0806627105}}.

\bibitem[\protect\citeauthoryear{Vijayan and
  Milenkovi{\'c}}{2017}]{Vijayan-2017-Multimagna}
V.~\textsc{Vijayan} and T.~\textsc{Milenkovi{\'c}}.
\newblock \href{http://dx.doi.org/10.1109/TCBB.2017.2740381}{\emph{Multiple
  network alignment via multimagna++}}.
\newblock IEEE/ACM Transactions on Computational Biology and Bioinformatics,
  PP~(99), pp. 1--1, 2017.
\newblock \href {http://dx.doi.org/10.1109/TCBB.2017.2740381}
  {\normalcolor\path{doi:10.1109/TCBB.2017.2740381}}.

\end{thebibliography}
\end{fullwidth}

\appendix
\section{The aposterori bound on the best single-rank alignment.}

We now show the result from the main text. Recall the setting:

We discuss a approaches to solve the $k$-dimensional matching problem: 
\begin{equation}
 \MAXone{}{\sum_{i,j,\ldots,\ell} \cT(i,j,\ldots,\ell) \cX(i,j,\ldots,\ell)}{\flat_1(\cmX) \ve \le \ve; \ldots; \flat_k(\cmX) \ve \le \ve; \cX(i,j,\ldots,\ell) \in \{0, 1\}}
\end{equation}
when $\cmT$ is given by a non-negative rank-$t$ representation: 
\begin{equation} \label{eq:low-rank-T}
\begin{aligned}
& \textstyle \cT(i,j,\ldots,\ell) = \sum_{t=1}^r U_1(i,t) U_2(j,t) \cdots U_k(\ell,t)  \\
& \textstyle \qquad \Leftrightarrow   \cmT = \sum_{i=1}^t \cmT_i \text{ where } \tvec(\cmT_i) = \mU_k(:,i) \kron \mU_{k-1}(:,i) \kron \cdots \kron \mU_1(:,i)
\end{aligned}
\end{equation}
that builds on an algorithm for low rank bipartite matchings from~\cite{Nassar-2018-spectral}.

Let the binary-valued tensors $\cmM_i$ of size $n_1 \times n_2 \times \hdots n_k$ store the matching corresponding to $\cmT_i$ tensor, i.e., $\cM(j_1,j_2,\hdots,j_k) = 1$ if $(j_1,j_2,\hdots,j_k)$ is a match, and $0$ otherwise.  By the generalized rearrangement inequality, these were optimal on the respective tensors. We now prove the following: 

\textbf{Result.} \emph{Consider the best $k$-dimensional matching from the set $\cmM_1, \ldots, \cmM_t$, then this is a $D$-approximation to the best $k$-dimensional matching, where $D$ is an aposterori computable bound.}

Define $\cmM_i \bullet \cmT_i = \tvec(\cmM_i)^T \tvec(\cmT_i)$ to be the weight of the matching $\cmM_i$ applied on the tensor $\cmT_i$. Also, let $\cmM^*$ to be the matching that achieves the maximum possible weight on $\cmT$. 

Define $d_{i,j} = \frac{\cmM_i \bullet \cmT_i }{\cmM_j \bullet \cmT_i}$, and $d_j = \max_i d_{i,j}$. Let $j^* = \text{argmin}_j d_j$. Set $D = d_{j^*}$. Then, $\cmM^* \bullet \cmT \leq D \cmM_{j^*} \bullet \cmT$. The proof of this statement follows: 
\[
\textstyle \cmM^* \bullet \cmT = \cmM^* \bullet \sum_{i=1}^{t} \cmT_i \leq \sum_{i = 1}^{t} \cmM_i \bullet \cmT_i \le D \sum_{i=1}^t (\cmM_{j^*} \bullet \cmT_i) \le D (\cmM_{j^*} \bullet \cmT),
\]
where we used $\cmM_i \bullet \cmT_i \le D \cmM_{j^*} \bullet \cmT_i$ by the definition of the quantities.
Therefore, the matching $\mM_{j^*}$ achieves a $D-$approximation on the tensor $\cmT$.

\section{Proof of the generalized rearrangement inequality.}
\label{sec:gen_rearrange}
Assume that we have $k$ sequences of numbers that are all positive. Let $x_i^{(j)}$ denote the $i^{th}$ element in the $j^{th}$ sequence and assume that $x_1^{(j)} \leq x_2^{(j)} \leq \hdots x_n^{(j)}$ for all sequences. The claim is:
\[
\sum_{i=1}^n \prod_{j=1}^{k} x_i^{(j)} \geq \sum_{i=1}^n x_i^{(1)} \prod_{j=2}^k x_{\sigma_j(i)}^{(j)}
\]
where $\sigma_j$ is any permutation function corresponding to the $j^{th}$ sequence. The proof follows a similar strategy as the proof of rearrangement inequality on two sequences and we extend it here. We prove this by induction. 

We first assume that we have $k$ sequences with $2$ elements each. We claim that
\[
\prod_{i=1}^{k} x_1^{(i)} + \prod_{i=1}^{k} x_2^{(i)}  \geq x_1^{(1)} \prod_{i=2}^{k} x_{\sigma_i{(1)}}^{(i)} + x_2^{(1)}\prod_{i=2}^{k} x_{\sigma_i{(2)}}^{(i)}
\]
We prove this by contradiction. Assume that there exists a permutation $\sigma$ such that the above formula is incorrect. Let us expand the right hand side.
\begin{align*}
x_1^{(1)} \prod_{i=2}^{k} x_{\sigma{(1)}}^{(i)} + x_2^{(1)}\prod_{i=2}^{k} x_{\sigma{(2)}}^{(i)} = 
x_1^{(1)} x_{\sigma_2{(1)}}^{(2)} \hdots x_{\sigma_k{(1)}}^{(k)}+ 
x_2^{(1)} x_{\sigma_2{(2)}}^{(2)} \hdots x_{\sigma_k{(2)}}^{(k)}
\end{align*}
Let $P_1 = x_1^{(1)}\prod x_{\sigma_j{(1)}}^{(j)} \forall j$ such that $\sigma_j(1) = 1$. Let $P_2 = \prod x_{\sigma_j{(1)}}^{(j)} \forall j$ such that $\sigma_j(1) = 2$. Similarly, let $Q_1 = \prod x_{\sigma_j{(2)}}^{(j)} \forall j$ such that $\sigma_j(2) = 1$, and $Q_2 = x_2^{(1)} \prod x_{\sigma_j{(2)}}^{(j)} \forall j$ such that $\sigma_j(2) = 2$. Then,
\[
x_1^{(1)} \prod_{i=2}^{k} x_{\sigma{(1)}}^{(i)} + x_2^{(1)}\prod_{i=2}^{k} x_{\sigma{(2)}}^{(i)} = P_1 P_2  + Q_1 Q_2
\]
Now observe that $P_2 \geq Q_1$ and $Q_2 \geq P_1$ by definition. Thus, using the rearrangement inequality on two sequences, $P_2 Q_2 + P_1 Q_1$ must be $\geq P_1 P_2 + Q_1  Q_2$, thus, a contradiction and the best way to arrange the terms in these sequences is by grouping all the bigger elements together and the smaller elements together.

Next, we assume that the property holds for $k$ sequences with $n-1$ elements each. We now prove it for $k$ sequences with $n$ elements each. We prove this by contradiction and assume that there exists $k-1$ permutations ($\sigma_2, \hdots \sigma_k)$ of size $n$ each which achieves the maximal pairing, i.e.
\[
\sum_{i = 1}^{n} x_i^{(1)} \prod_{j=1}^{k} x_{\sigma_j{(i)}}^{(j)}
\]
If this sum is the maximal sum, then there are two terms in this summation of the form:
\[
x_1^{(1)} x_{i_2}^{(2)} x_{i_3}^{(3)} \hdots x_{i_k}^{(k)} + x_{j_1}^{(1)} x_{1}^{(2)} x_{j_3}^{(3)} \hdots x_{j_k}^{(k)}
\]
We can now apply the base case on the following $k$ sequences with 2 elements:
\[
\{x_1^{(1)}, x_{j_1}^{(1)}\},
\{x_1^{(2)}, x_{i_2}^{(2)}\},
\{x_{i_3}^{(3)},x_{j_3}^{(3)}\}, \hdots,
\{x_{i_k}^{(3)},x_{j_k}^{(3)}\}
\]
Without loss of generality, assume that $x_{i_l}^{(l)} \leq x_{j_l}^{(l)}$ for all $ 3\leq l \leq k$. Thus, by using the base case, we know that $x_1^{(1)} x_1^{(2)} x_{i_3}^{(3)} \hdots x_{i_k}^{(k)} + x_{j_1}^{(1)} x_{i_2}^{(2)} x_{j_3}^{(3)} \hdots x_{j_k}^{(k)} \geq x_1^{(1)} x_{i_2}^{(2)} x_{i_3}^{(3)} \hdots x_{i_k}^{(k)} + x_{j_1}^{(1)} x_{1}^{(2)} x_{j_3}^{(3)} \hdots x_{j_k}^{(k)}$. This means that we have just found a rearrangement of two terms in the maximal summation that can be rearranged to achieve a higher weight, which is a contradiction. We proceed with the same strategy to show that the permutations should indeed be the identity permutations. Next, we consider the following two terms:
\[
x_1^{(1)} x_{1}^{(2)} x_{i_3}^{(3)} \hdots x_{i_k}^{(k)} + x_{j_1}^{(1)} x_{j_2}^{(2)} x_{1}^{(3)} \hdots x_{j_k}^{(k)}
\]
Similarly, and assuming that $i_l \leq j_l$, we can conclude that
\[x_1^{(1)} x_{1}^{(2)} x_{1}^{(3)} \hdots x_{i_k}^{(k)} + x_{j_1}^{(1)} x_{j_2}^{(2)} x_{j_3}^{(3)} \hdots x_{j_k}^{(k)} \geq x_1^{(1)} x_{1}^{(2)} x_{i_3}^{(3)} \hdots x_{i_k}^{(k)} + x_{j_1}^{(1)} x_{j_2}^{(2)} x_{1}^{(3)} \hdots x_{j_k}^{(k)}.
\]
If we proceed with the same strategy for all the remaining terms, we will achieve a summation of the form
\[
x_1^{(1)} x_1^{(2)} \hdots x_1^{(k)} + \sum_{i = 2}^{n} x_i^{(1)} \prod_{j=1}^{k} x_{\sigma_j{(i)}}^{(j)}
\]

By the inductive hypothesis, we know that $\sum_{i = 2}^{n} x_i^{(1)} \prod_{j=1}^{k} x_{i}^{(j)} \geq \sum_{i = 2}^{n} x_i^{(1)} \prod_{j=1}^{k} x_{\sigma_j{(i)}}^{(j)}$, and hence each of the permutations must be the identity permutation.
\[
\sum_{i = 1}^{n} x_i^{(1)} \prod_{j=1}^{k} x_{\sigma_j{(i)}}^{(j)}
\]
And with that, the rearrangement inequality for k sequences of size n each is proved. For cases when the number of elements in each sequence is different, we set $n$ to be the size of the smallest sequence and pick the top $n$ elements of each of the other sequences. The reason we pick the top $n$ can be viewed as a direct application of the rearrangement inequality.

\section{Eight iterations is enough}
\label{sec:eight_iters}
\begin{marginfigure}
	\centering
		\centering \includegraphics[width=\linewidth]{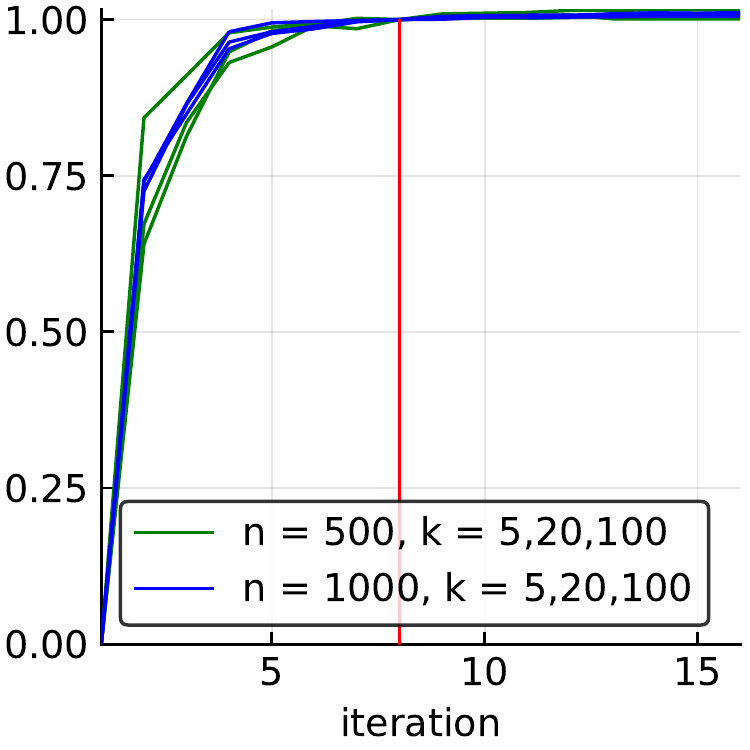}
		\footnotesize \emph{Preferential attachment}
		\centering \includegraphics[width=\linewidth]{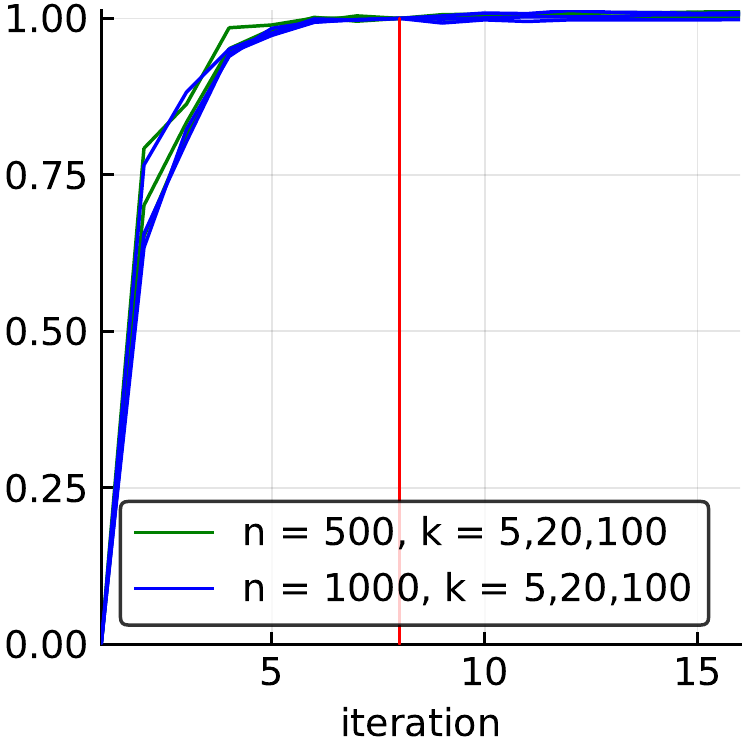}
		\caption{To make sure that 8 iterations of the power method are enough to achieve a good result, we run our algorithm MultiLR-D for other iteration values and discover that after 8 iterations essentially nothing changes. The y-axis in these plots is the degree weighted recovery value relative to the value at iteration 8; there are 6 curves for three different settings in terms of the number of networks and the size of the networks that are all indistinguishable.}
		\label{fig:iters}
	\end{marginfigure}

Here we show a case study on running our algorithm on several kinds of problems with varying $k$ and varying $n$. We plot the degree weighted recovery normalized by the value on iteration 8. These results show that the quality of the result does not change considerably after iteration 8 (see Figure~\ref{fig:iters}).
	
\section{The values of D are small in practice}
\begin{figure}
		\begin{minipage}{0.66\linewidth}
		\centering \includegraphics[width=\linewidth]{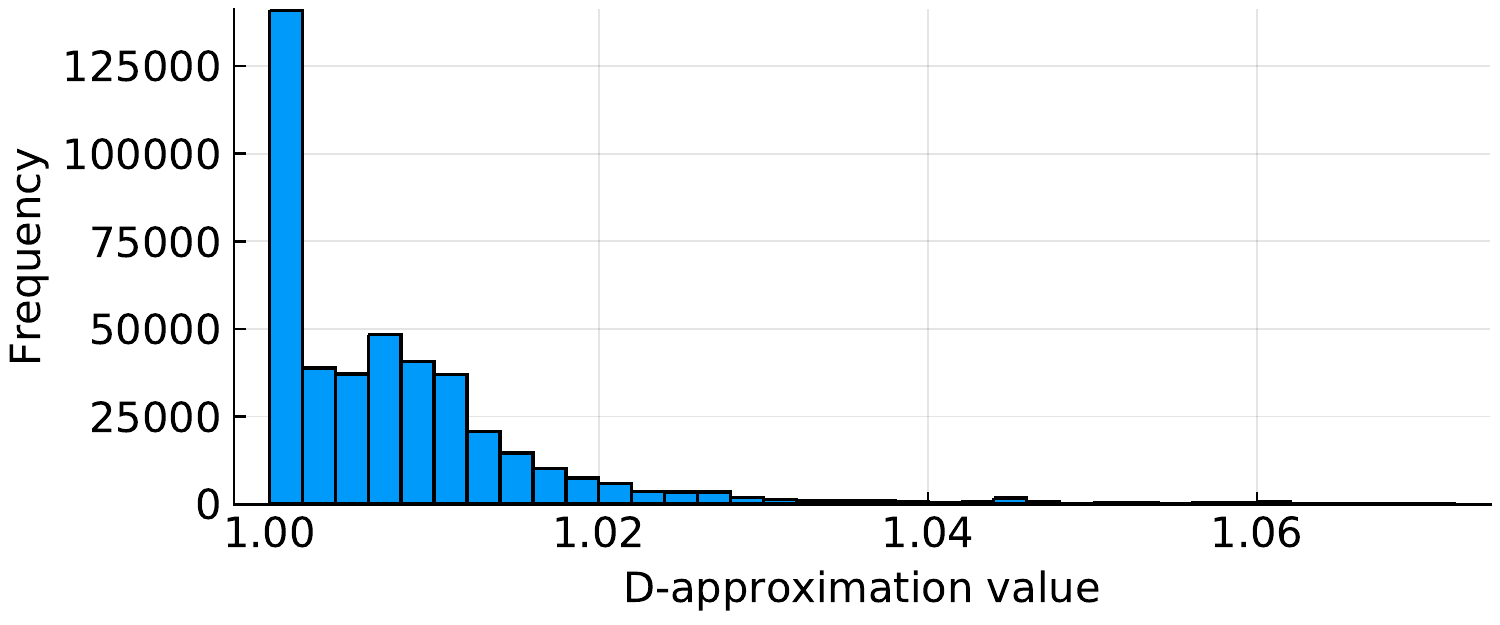}
		\end{minipage}
		\begin{minipage}{0.33\linewidth}
		\caption{This figure shows the D approximation values from the experiment of aligning multiple egonets in the DBLP network. These numbers show that the approximation bound D is very close to $1$ in practice.}
		\end{minipage}
		\label{fig:dblp_histogram}
	\end{figure}

In the experiment aligning the egonets from the DBLP network, we only ran our MultiLR-D, and here we show the quality of the posterior bound $D$ we get from this method. We plot a histogram of these values here (figure~\ref{fig:dblp_histogram}) and observe that a striking number of them is very close to $1$, and even the maximum of them is still less than $<1.1$. For other cases when we ran our MulitLR-D algorithm, the values were comparable as well with the maximum less than $1.07$ or $1.08$. 

\section{The synthetic ER and PA graph constructions}
\label{sec:graph_construction}
For ER, we set the edge probability such that we achieve the expected degree $d$ and $n$ nodes. For PA, to generate a graph with $n$ edges, we start with a 5-node clique graph and add $\theta$ edges from each new vertex following the preferential attachment model. The expected degree is $2 \theta$ because each new edge gets counted twice in the average degree computation. 
 
Then to generate $k$ instances of these graphs, we generate one reference graph, and then we then pick an edge deletion probability $p_e$, and generate $k$ instances of the base graph, we allow an each edge to be deleted according the the probability $p_e$. We repeat this process $k$ times to reach $k$ networks.

\section{More alignment baselines}
\label{sec:additional_algs}
\paragraph{Progressive EigenAlign} We mention the recent pairwise network alignment algorithm EigenAlign in~\cite{Feizi-2016-spectral}, and its low rank formulation from~\cite{Nassar-2018-spectral} to be strong pairwise network alignment algorithm when no prior information about node similarity is present. Here, we suggest a simple extension to this algorithm to adapt it to align multiple networks and we follow a progressive approach. We start with two networks to align them using the low rank formulation of EigenAlign from~\cite{Nassar-2018-spectral}. After the first two networks are aligned, we fold them on top of each other by using the new matches to form a new network. Then, we use this network to align it to the next network. For $k$ networks, the pairwise procedure would occur $k-1$ times.
\paragraph{Random} Another method we choose to compare our existing methods to is a random alignment. This is more of a sanity check experiment to make sure that the algorithms we are using do not generate arbitrary matchings and that indeed a random matching would not outperform any of the existing methods.

\section{More on the routers dataset}
Here, we show the same results as in the routers section from the main text, adding to it the two new methods described here (Progressive EigenAlign as well as random). From figure~\ref{fig:routers_all}, we see that Progressive EigenAlign was giving good results, and this is due to it being a strong pairwise method, and also due to the problem having only 5 networks. In the following section we study Progressive EigenAlign further on synthetic graphs as we vary the number of networks to be aligned. 
\begin{tuftefigure}
		\begin{minipage}{\linewidth}
		\centering \includegraphics[width=\linewidth]{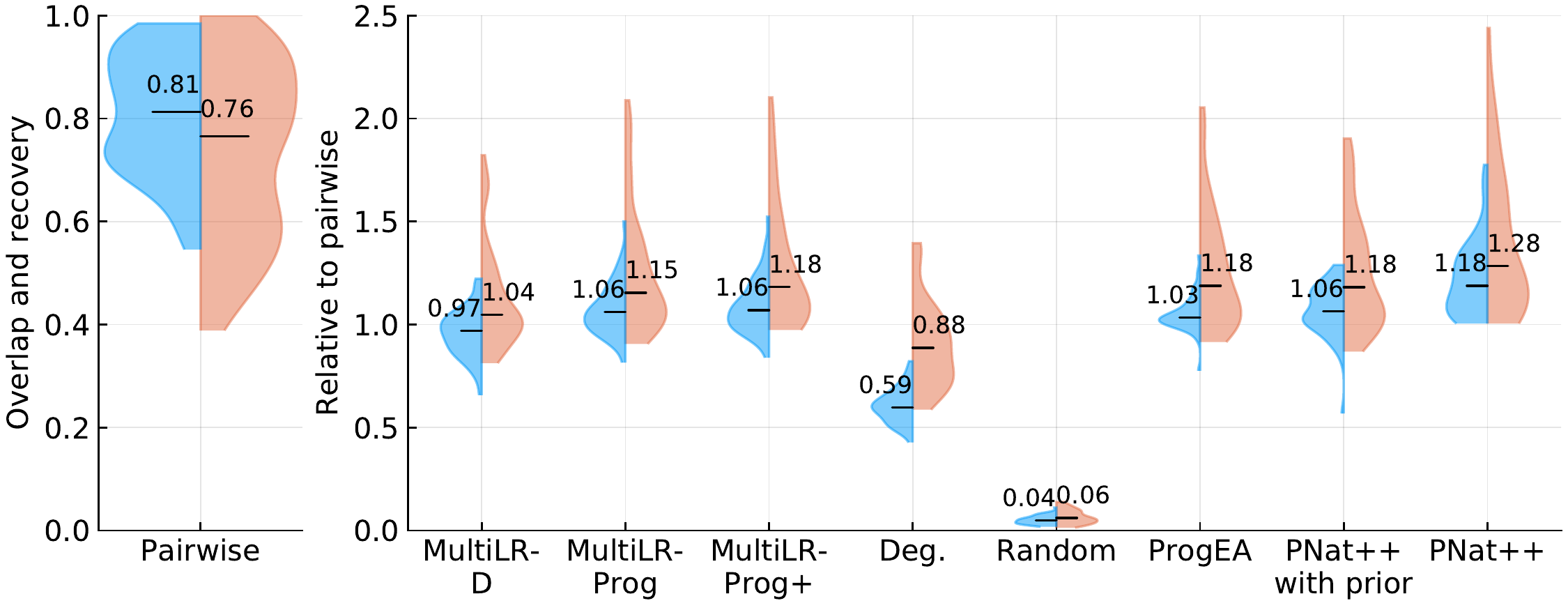}
		\footnotesize \emph{Top 25 nodes}
		\centering \includegraphics[width=\linewidth]{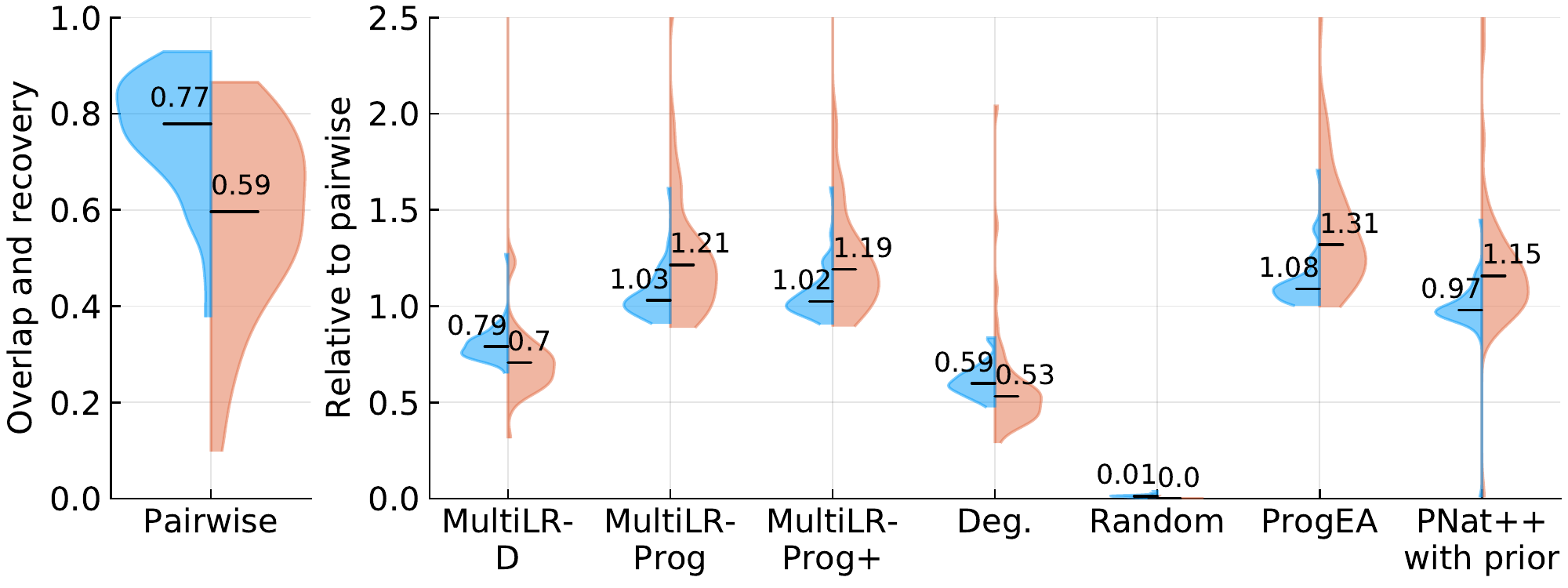}
		\footnotesize \emph{Top 100 nodes}
		\end{minipage}
		\caption{In this figure, we add the results from Progressive EigenAlign (ProgEA in the plot) as well as the random method. We observe that in both experimental settings, we always perform better than the random method.}
		\label{fig:routers_all}
\end{tuftefigure}

\section{Progressive EigenAlign fails with large numbers of networks}
In Figure~\ref{fig:er_vs_pa_all} we replot a figure from the main text with Progressive EigenAlign added. This method performs poorly with large numbers of networks. 

\begin{fullwidthfigure}
\parbox{0.5\linewidth}{\centering \itshape As edge deletion varies \ldots }%
\parbox{0.5\linewidth}{\centering \itshape As we add networks \ldots }
\parbox[c]{0.25\linewidth}{\centering \footnotesize ~~~~Erd\H{o}s-R\'enyi}%
\parbox[c]{0.25\linewidth}{\centering \footnotesize ~~PA}%
\parbox[c]{0.25\linewidth}{\centering \footnotesize ~~~~Erd\H{o}s-R\'enyi}%
\parbox[c]{0.25\linewidth}{\centering \footnotesize ~~PA}
\includegraphics[width=0.25\linewidth]{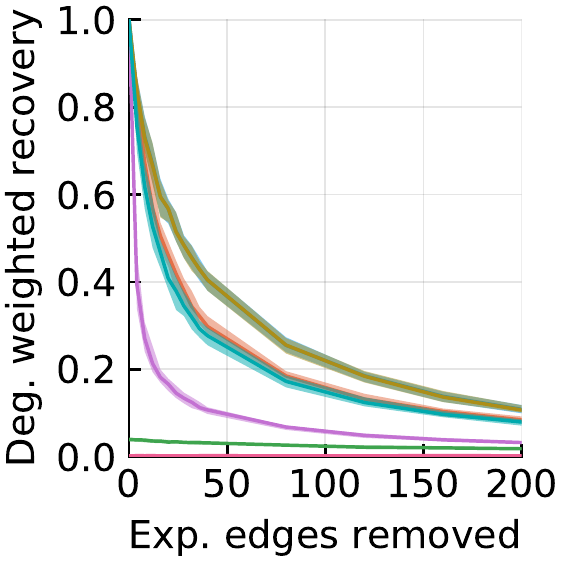}%
\includegraphics[width=0.25\linewidth]{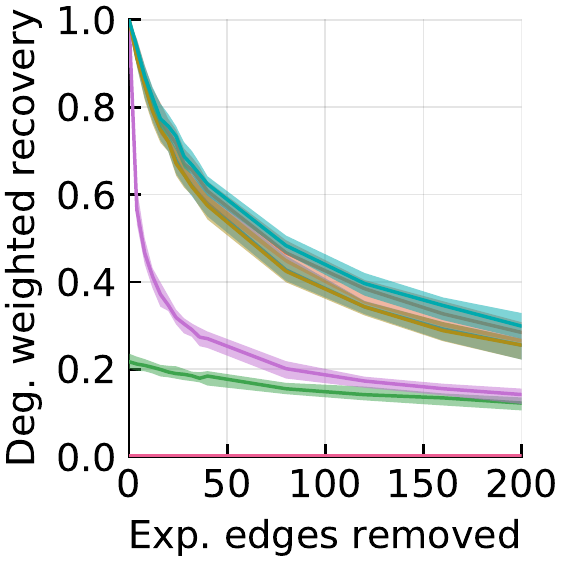}%
\includegraphics[width=0.25\linewidth]{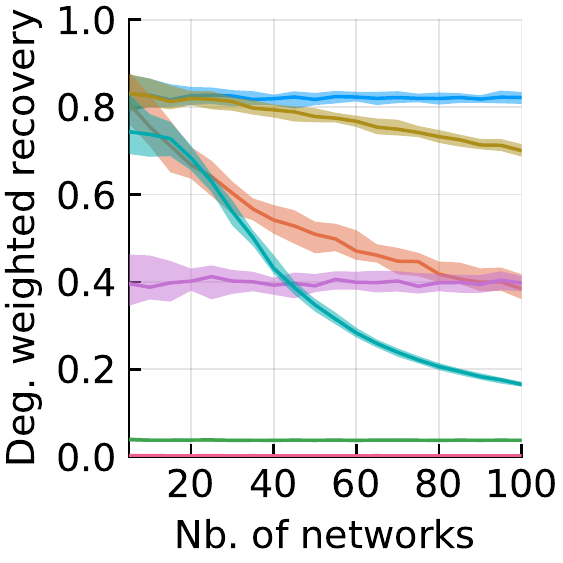}%
\includegraphics[width=0.25\linewidth]{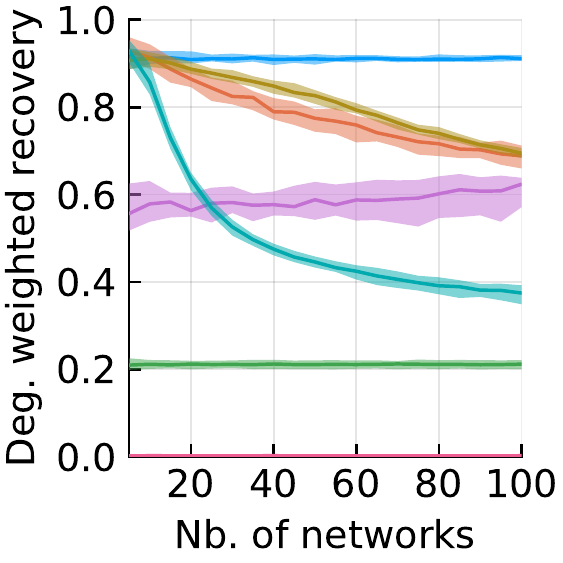}

\parbox{\linewidth}{\centering \footnotesize
\legendbox{plots1}
MultiLR-Prog+ $\quad$ 
\legendbox{plots2}
Pairwise $\quad$ 
\legendbox{plots3}
Degree $\quad$ 
\legendbox{plots4}
MultiLR-D $\quad$ 
\legendbox{plots5}
MultiLR-Prog $\quad$ \\
\legendbox{plots6}
ProgEigenAlign $\quad$ 
\legendbox{plots7}
Random $\quad$ 
}
\caption{An extended version of a plots from the main paper with both random and Progressive EigenAlign methods added. This shows Progressive EigenAlign's quality degrades with large numbers of networks. }
\label{fig:er_vs_pa_all}
\end{fullwidthfigure}

\section{As the network size varies.}
In this experiment, we are interested in observing how the alignment quality varies as we change the sizes of the networks to be aligned. Here, we use preferential attachment graphs and we fix the edge deletion probability to $p_e = 0.5/n$, as we vary $n$. We observe that all methods are essentially resistant to the change in the network sizes whereas this behavior is not true when the number of networks become much bigger (such as 100). From figure~\ref{fig:prefattach_varyn}, we can conclude that MultiLR-Prog+ is resistant to both changes in the network sizes, as well as the number of networks to be aligned. Interestingly, MultiLR-D is also resistant to such changes but with a worse recovery score.
\begin{tuftefigure}
\parbox{\linewidth}{\centering
\legendbox{plots1}
MultiLR-Prog+ $\quad$ 
\legendbox{plots2}
Pairwise $\quad$ 
\legendbox{plots3}
Degree $\quad$ 
\legendbox{plots4}
MultiLR-D $\quad$\\
\legendbox{plots5}
MultiLR-Prog $\quad$ 
\legendbox{plots6}
ProgEigenAlign $\quad$ 
\legendbox{plots7}
Random $\quad$ 
}
\centering
	\begin{minipage}{0.33\linewidth}
		\centering \includegraphics[width=\linewidth]{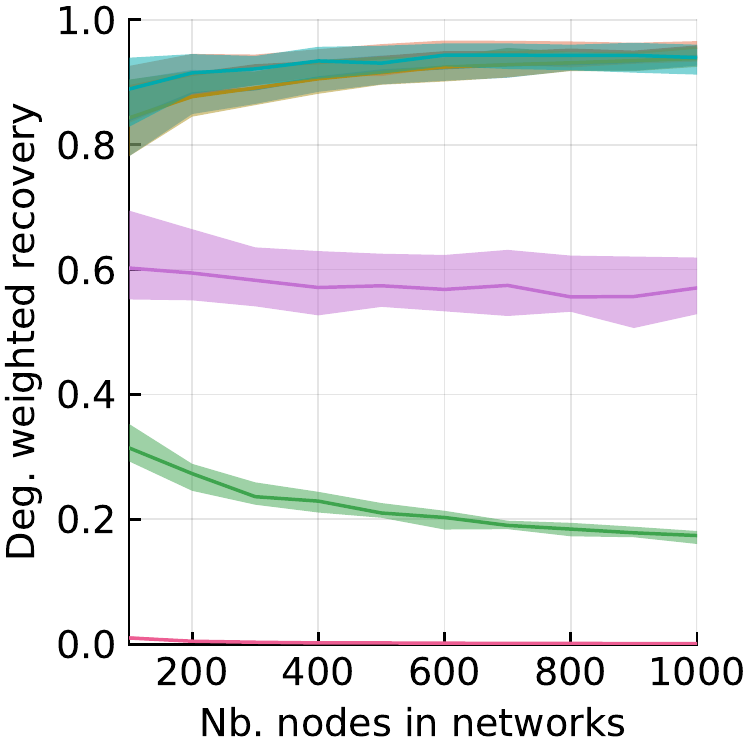}\\
		\footnotesize \emph{(a): k = 5 networks}
	\end{minipage}%
	\begin{minipage}{0.33\linewidth}
		\centering \includegraphics[width=\linewidth]{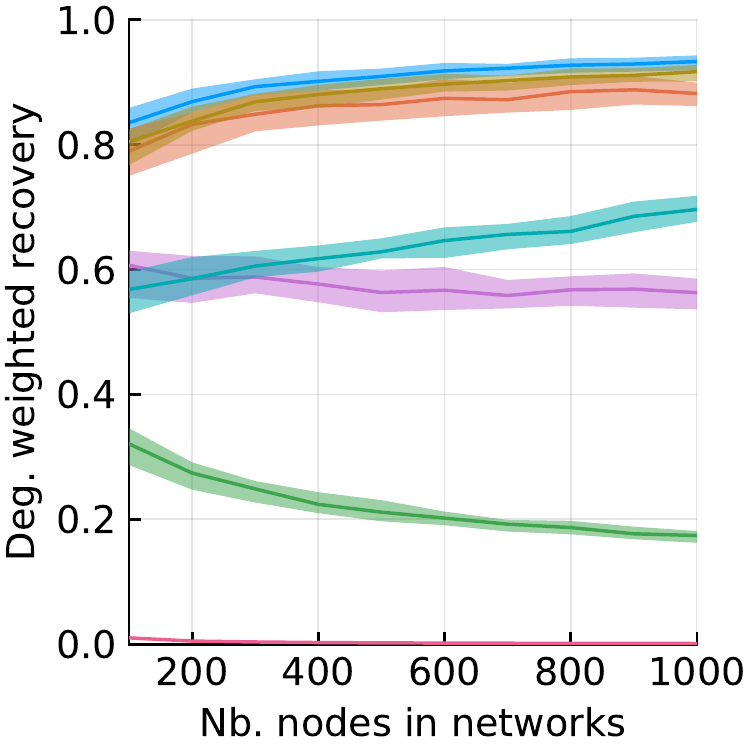}\\
		\footnotesize \emph{(b): k = 20 networks}
	\end{minipage}%
	\begin{minipage}{0.33\linewidth}
		\centering \includegraphics[width=\linewidth]{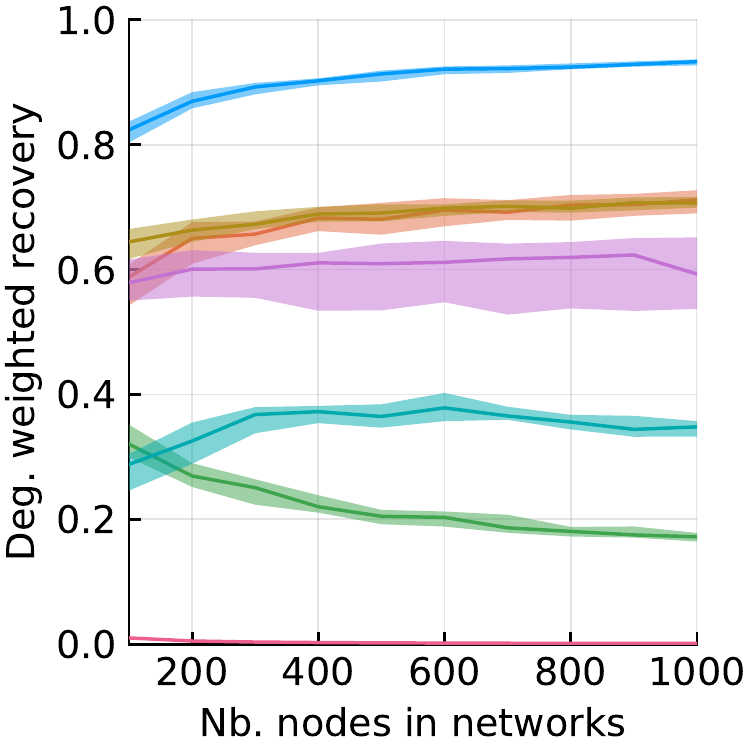}\\
		\footnotesize \emph{(c): k = 100 networks}
	\end{minipage}
	\vspace{-0.5\baselineskip}
	\caption{These figures show the weighted recovery scores on preferential attachment graphs as we vary the sizes of the networks, and the number of networks to be aligned. We observe that when the number of networks is small enough (5 networks) pairwise and multiple alignment methods achieve similar results. Whereas when we increase the number of networks to be aligned, multiple alignment sustains its result whereas the pairwise method  fails to do so.}
\label{fig:prefattach_varyn}
\end{tuftefigure}
\vspace{-1\baselineskip}
\section{Runtime information}
\label{sec:runtime}
Finally, in Figure~\ref{fig:table} we show runtime information for MultiLR-D and MultiLR-Prog+ for synthetic networks as we increase size and the number of networks up to thousands. Then in Table~\ref{tab:times25} and Table~\ref{tab:times100} we show runtimes for the methods on the routers alignment problems.
\begin{tuftefigure}
	\begin{minipage}{0.5\linewidth}
		\centering \includegraphics[width=\linewidth]{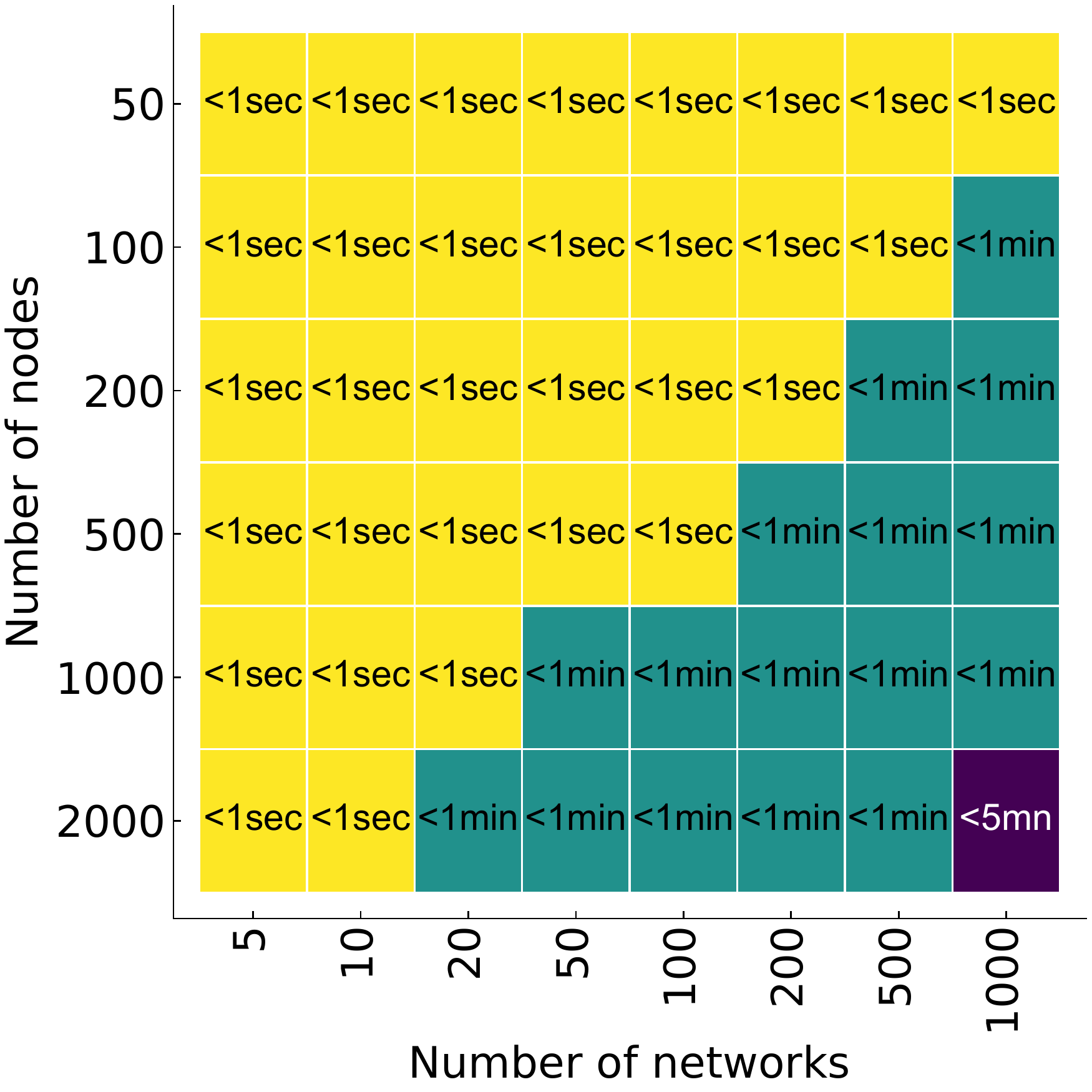}
		\footnotesize \emph{MultiLR-D}
	\end{minipage}
	\begin{minipage}{0.5\linewidth}
		\centering \includegraphics[width=\linewidth]{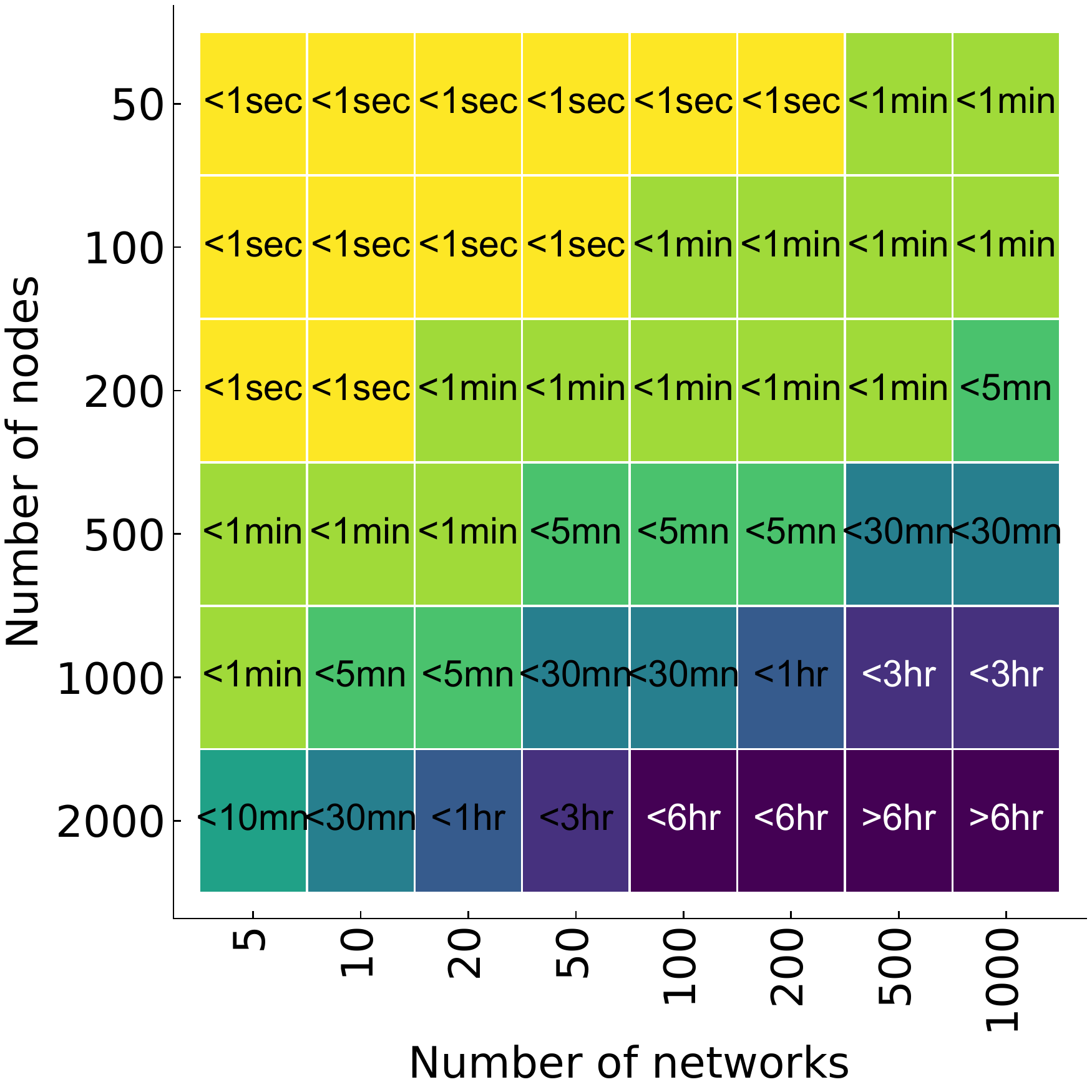}
		\footnotesize \emph{MultiLR-Prog+}
	\end{minipage}
	\caption{The runtime as we run MultiLR-D and MultiLR-Prog+ on the synthetic experiments on a wide variety of problem sizes with \erdosrenyi graphs.  }
	\label{fig:table}
\end{tuftefigure}
\begin{table}
\caption{Runtimes for the 25-node $5$-network router problem}\vspace{-0.5\baselineskip}
	\label{tab:times25}
	\begin{tabularx}{\linewidth}{lXXX}
	\toprule
	\multicolumn{1}{l}{Algorithm} 	& \multicolumn{3}{l}{Time (sec)} \\
			                               		&  min & max & median\\ \midrule
			
             MultiLR-D         				& 0.268854&   	0.406164&    	0.279505\\
             MultiLR-Prog              		& 0.354238&   	0.481346&    	0.375237\\
             MultiLR-Prog+				&0.305996 &     0.515386&     0.329824 \\
			Degree                   			&0.0229719&   	0.0436085&   0.024294\\
			Random						&0.0131963&     0.024009&     0.0143561\\
			ProgEigenAlig        		    &1.33297&     	1.48824&     	1.39133\\
			Pairwise						&5.51146&     	6.2133&       	5.8608\\
			Prognatalie++ with prior 	&2.17744&   		241.116&       23.095\\
			Prognatalie++                	&313.088&    		2823.43&       852.095\\  \bottomrule
		\end{tabularx}
\caption{Runtimes for the $100$-node $5$-network router problem}\vspace{-0.5\baselineskip}
	\label{tab:times100}
	\begin{tabularx}{\linewidth}{lXXX}
	\toprule
	\multicolumn{1}{l}{Algorithm} 	& \multicolumn{3}{l}{Time (sec)} \\
			                               		&  min & max & median \\ \midrule
			
             MultiLR-D         				& 0.356948&      0.465534&     0.371156\\
             MultiLR-Prog              		&0.321636&      0.501205&     0.348855\\
             MultiLR-Prog+				&0.364909&      0.484608&     0.384219 \\
             Degree                   			& 0.0234081&     0.0473191&    0.0249071\\
             Random						&0.0138897&     0.0262094&    0.0151449\\
			ProgEigenAlign        		&2.31544&       2.59832&      2.44014\\
			Pairwise						&4.88272&       5.37912&      5.04676\\
			Prognatalie++ with prior 	&86.0972&     1451.2&        649.47\\ \bottomrule
		\end{tabularx}
\end{table}
\end{document}